\documentclass[twoside]{article}

\usepackage{epsfig}

\usepackage{rotating}	

\usepackage{textcomp}

\newcommand{\tomreg}{\emph{De\-mand\_re\-gion}}
\newcommand{\hankreg}{\emph{Phas\-ed\_re\-gion}}

\newcommand{\etal}{et~al.}
\newcommand{\bi}{\begin{itemize}}
\newcommand{\ei}{\end{itemize}}
\newcommand{\be}{\begin{enumerate}}
\newcommand{\ee}{\end{enumerate}}

\newcommand{\bheading}[1]{\noindent\textbf{#1}}

\newcommand{\nicelooking}{\frenchspacing\widowpenalty=10000\clubpenalty=10000}

\newcommand{\Hzero}{\textbf{H\O}}
\newcommand{\Hone}{\textbf{H1}}
\newcommand{\Htwo}{\textbf{H2}}
\newcommand{\Hthree}{\textbf{H3}}
\newcommand{\Hfour}{\textbf{H4}}
\newcommand{\Hfive}{\textbf{H5}}
\newcommand{\Hsix}{\textbf{H6}}

\def\numbername#1{\ifcase#1%
   zero%
   \or one%
   \or two%
   \or three%
   \or four%
   \or five%
   \or six%
   \or seven%
   \or eight%
   \or nine%
   \or ten%
   \or #1%
   \fi
}%

\newcommand{\tabtriplecode}[3]{
 \hrule\vspace{6pt}
 \begin{tabular}{@{\hspace{10pt}}l@{\hspace{10pt}}@{\vline}@{\hspace{10pt}}l@{\hspace{10pt}}@{\vline}@{\hspace{10pt}}l}
 \begin{minipage}[t]{0.334\textwidth}
 \scriptsize
 \begin{tabbing}
 #1
 \end{tabbing}
 \end{minipage}
 &
 \begin{minipage}[t]{0.333\textwidth}
 \scriptsize
 \begin{tabbing}
 #2
 \end{tabbing}
 \end{minipage}
 &
 \begin{minipage}[t]{0.333\textwidth}
 \scriptsize
 \begin{tabbing}
 #3
 \end{tabbing}
 \end{minipage}
 \end{tabular}
 \vspace{6pt}\hrule
}

\title{Demand-driven Inlining in a Region-based Optimizer for ILP Architectures}
\author{
Thomas P. Way \\
Department of Computing Sciences \\
Villanova University, Villanova, PA 19085 \\
thomas.way@villanova.edu \\[10pt]
Lori L. Pollock \\
Department of Computer \& Information Sciences \\
University of Delaware, Newark, DE 19716 \\
pollock@cis.udel.edu}

\begin{document}
\maketitle

\nicelooking

\begin{abstract}

Region-based compilation repartitions a program into more 
desirable compilation units using profiling information
and procedure inlining to enable region formation analysis.
 Heuristics play a key role in determining 
when it is most beneficial to inline procedures
during region formation.
An ILP optimizing compiler using a region-based approach
restructures a program to better reflect
dynamic behavior and increase interprocedural optimization
and scheduling opportunities.
This paper presents an interprocedural compilation technique
which performs procedure inlining on-demand, 
rather than as a separate phase,
to improve the ability of a region-based optimizer to
control code growth, compilation time and memory usage
while improving performance.
The interprocedural region formation algorithm 
utilizes a demand-driven, heuristics-guided 
approach to inlining, restructuring an input program into 
interprocedural regions.
Experimental results are presented to demonstrate
the impact of the algorithm and several  
inlining heuristics upon a number of traditional and novel 
compilation characteristics within a region-based ILP 
compiler and simulator.

\end{abstract}

\section{Introduction}
\label{section:intro}

Advanced instruction-level parallel (ILP) computer architectures
require aggressive and potentially costly whole program, 
or interprocedural, techniques for program analysis and optimization 
to fully exploit available parallelism.
These interprocedural techniques are in contrast to  
intraprocedural code improvement techniques employed
in a traditional procedure-oriented compiler,
where analysis and optimization
phases 
are independently applied to each 
procedure in isolation.~\cite{Aho88}

An approach for ILP that reduces the cost of aggressive
interprocedural analysis and optimization is region-based
compilation~\cite{Hank97}.
Region-based compilation is a generalized
trace selection approach that partitions
a program into units of compilation, or \emph{regions},
based on profile information.
Using procedure inlining, where a procedure callsite is 
replaced by the body of the called procedure,
and restructuring a program into regions,
the region-based compiler can
perform code motion and other analyses and optimizations
interprocedurally, while maintaining control
over the compilation unit size and content.
Unlike procedure-based compilation,
region-based techniques bound the compilation unit size
to better control optimization costs~\cite{Hank97}.

The key component of a region-based compiler is the 
region formation phase which partitions the program into 
regions using profile-guided heuristics with the intent
that the ILP optimizer will be invoked with a scope that is limited to
a single region at a time.
Thus, the quality of the generated code depends greatly upon the 
ability of the region formation phase to create regions that
a global optimizer can effectively transform in isolation for improved ILP.
Because region-based compilation relies on an initial 
aggressive inlining phase, region formation
remains quite costly,
particularly for large programs with many procedures and calls~\cite{Hank97}.
Selective use of inlining can
prevent excessive code growth and control register pressure
while improving analysis opportunities and performance~\cite{Ayer97}.

In this paper, a strategy to overcome the issues caused by separate 
inlining and region formation phases is described and evaluated. 
Presented is a demand-driven approach to inlining and a set of
inlining heuristics which are integrated within a 
region-based optimizer.
To evaluate these techniques, the algorithm and various
heuristics for guiding inlining decisions have been implemented
within the Trimaran ILP research compiler~\cite{Trim98}.
In addition to standard metrics such as compilation
time, code growth and execution time,
novel metrics have been devised to compare the characteristics of 
regions, such as profile homogeneity and interprocedural scope,
to measure the effectiveness of this new approach.

\section{Region-based Compilation}
\label{section:background}

A common characteristic of compiler
analysis techniques, including those specifically
for ILP architectures, is that they have
been designed with the assumption that the original
procedure boundaries created by the programmer are immutable.
Procedures serve as the \emph{de facto} unit of compilation.
As a result, there is the potential for large procedures
to either unacceptably increase compilation time or to
be less aggressively optimized (or not optimized at all) in
order to control compilation costs and maintain scalability.  
Procedure boundaries are a natural impediment to 
compilation effectiveness in many cases, 
requiring tradeoffs in terms of quality of
optimization versus compilation time and memory requirements.

Hank \etal~\cite{Hank97} proposed the region-based compilation
framework as a solution to the problem of exposing interprocedural
scheduling and optimization opportunities without the cost of
very large procedure bodies created through inlining, or the expense and
complexity of sophisticated interprocedural analysis and code motion. 
While it was shown to be especially beneficial in an ILP compiler,
region-based compilation also can achieve both
interprocedural scope and scalability in program analysis.

\subsection{Fundamental region formation}

\begin{figure}
\centering
\begin{tabular}{c}
\epsfig{figure=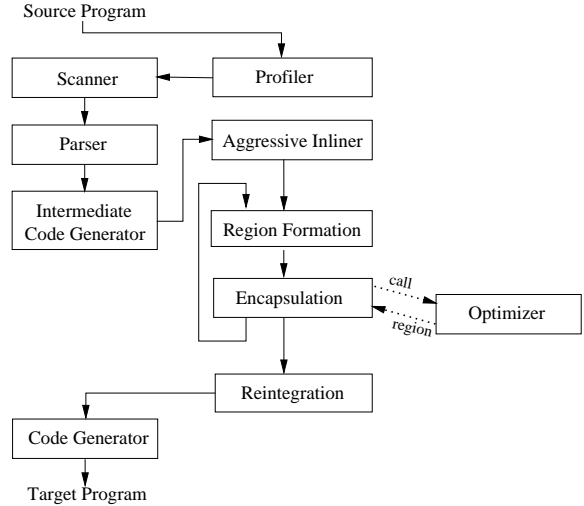,width=3.0in} \\     
\end{tabular}
\caption{Original region-based compilation framework.~\cite{Hank97}}
\label{fig:frameworks-hank}
\end{figure}

Figure~\ref{fig:frameworks-hank} depicts the
organization of a region-based compiler framework.
The source code enters the \emph{Profiler}, where the source
code is instrumented and executed to gather profile
information which is then integrated into the source code. 
Intermediate code with profiling information
is input to the \emph{Aggressive Inliner} phase, 
where all inlining that can be done
in the entire program, subject to some
constraints, is performed.
Next, in the \emph{Region Formation} phase,
regions are formed throughout the whole program,
and each region is encapsulated as a procedure in
the \emph{Encapsulation} phase.
The encapsulated regions are then passed to
a high-level \emph{Optimizer} phase before \emph{Reintegration}
into their original procedures.
The result is passed to the \emph{Code Generator} which
includes a low-level optimization phase.

In this framework, a \emph{region} is a collection of
basic blocks and control flow edges selected for
compilation as a unit~\cite{Hank97}.
More formally, a region is a subgraph of the 
control flow graph (CFG) of a procedure, created either
based on the structure of the CFG or using profile information. 
Each region is encapsulated in a 
single-entry, single-exit CFG by adding dummy
prologue and epilogue CFG nodes and boundary condition 
CFG nodes that convey pertinent data flow information.
Regions are encapsulated in such a way that
the optimizer can be invoked with a scope that is limited 
to a given region, which then appears to the rest of the compiler as a procedure.
Side entries into regions can be removed by tail duplication, similar
to superblock formation~\cite{Hwu93}.
After optimization, each region is reintegrated into the
original procedure in which the region existed by updating 
changes in data flow conditions, 
entry and exit points, and constraints on register allocation.
Code is generated from the reintegrated procedure.

\subsection{Example}

\begin{figure}
\centering
\small
\begin{tabular}{c@{\hspace{0.35in}}c@{\hspace{0.35in}}c}
\epsfig{figure=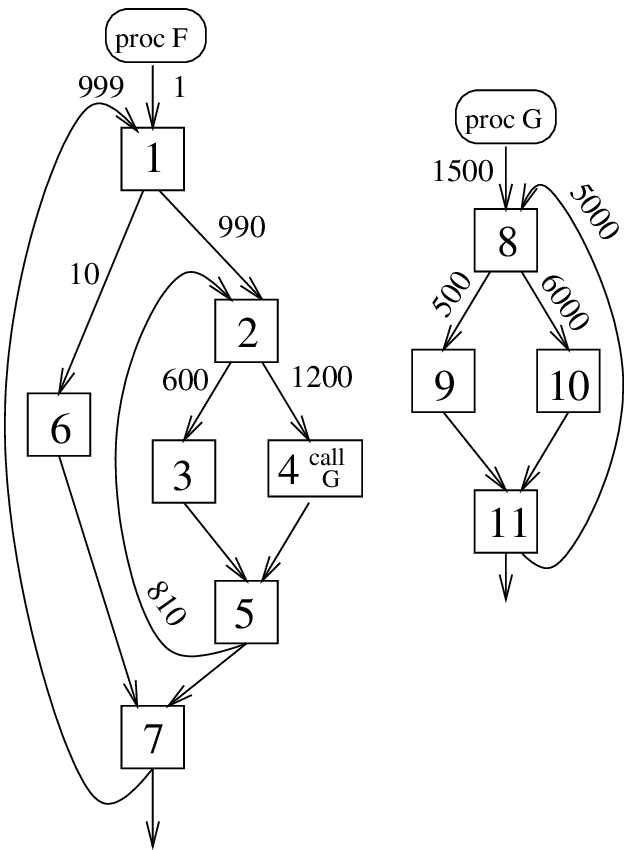,width=1.8in} &    
\epsfig{figure=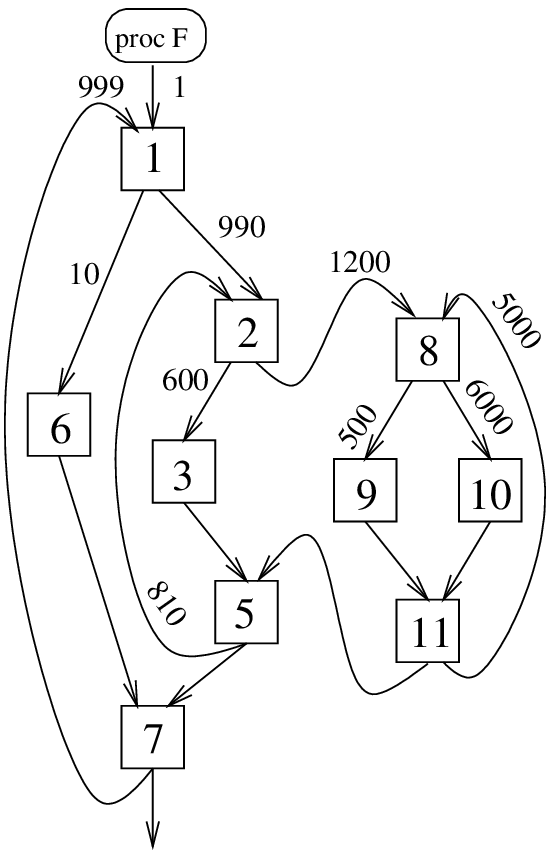,width=1.9in} &
\epsfig{figure=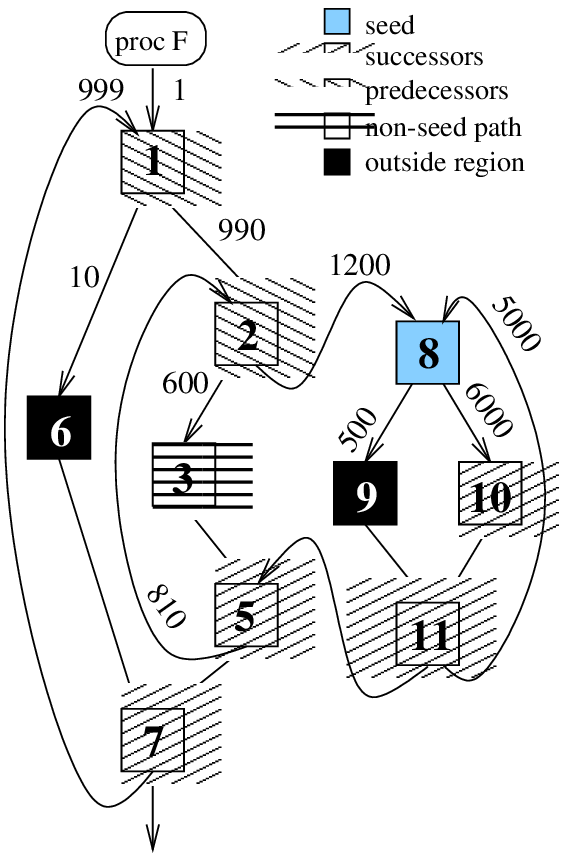,width=1.8in} \\
\multicolumn{1}{p{1.8in}}{\textbf{(a)}
     Procedures F and G prior to aggressive inlining.} &
\multicolumn{1}{p{1.9in}}{\textbf{(b)}
     Procedure G is inlined into F.} &
\multicolumn{1}{p{1.8in}}{\textbf{(c)}
     Region formation is performed in new F.} \\
& \\[6pt]
\end{tabular}
\normalsize
\caption{Example of the steps in Hank's region formation algorithm.}
\label{fig:4steps}
\end{figure}

The original profile-sensitive region formation algorithm
is comprised of the following steps, performed between
aggressive inlining and region encapsulation. 
These steps are performed until all blocks in the program have
been included in some region.
Figure~\ref{fig:4steps} shows the results of performing
the following steps of the algorithm.
Figure~\ref{fig:4steps}(b) shows the code after aggressive
inlining is performed on the code in Figure~\ref{fig:4steps}(a).

%
%

\bheading{Step 1: Seed Selection} - From among all basic
blocks in the procedure
not yet included in a region, select
the block with the highest execution frequency; this is the seed block for a new region.
In this simplified example, this is block 8, shown in Figure~\ref{fig:4steps}(b).
Note that inlining was done previously.

\bheading{Step 2: Region Expansion to Successors} - 
A path of \emph{desirable} successors is selected, 
starting at the seed block.
Region expansion is guided by heuristics which halt
the growth under a set of conditions such as~\cite{Hank97}:
(1) a procedure call is reached,
(2) a minimum acceptable execution frequency 
for a successor block is not met (e.g., at least 50\%
of the frequency of both its immediate predecessor
in the region and that of the seed block, 
which in this simplified example is why 
block 6 is not selected in this step), or
(3) a region size threshold (e.g., 200 basic blocks) 
is exceeded. The successors selected for seed block 8 
are blocks 10, 11, 5 and 7.


\bheading{Step 3: Region Expansion to Predecessors} - 
A path of frequently executed predecessors 
to the seed block is chosen
analogous to the selection of desirable successors.
The resulting path after this step is the seed path of the region.
In this case, blocks 2 and then 1 are added as predecessors of seed block 8.

\bheading{Step 4: Region Expansion from All Blocks in the Seed Path} - 
By selecting as above the desirable successors of \emph{all} 
current blocks in the region, 
the region is grown along multiple control flow paths.  
Thus, block 3 is added to the region.
The result of this step is a path-sensitive region.
Blocks not yet in a region (blocks 6 and 9) 
are used to form additional regions.


%

%

To summarize, three regions are formed in the example.
The largest region consists of blocks 1, 2, 3, 5, 7, 8, 10, and 11.
The remaining blocks 6 and 9 form single block regions.  
Note that original block 4 was replaced by the inlined procedure G.
%
%
Limitations include the potential for excessive code growth and 
unnecessary inlining due to the aggressive approach to inlining,
leading to unscalability,
and the training-data effect of profile-guided compilation.  
While Hank's approach can achieve scalability during program analysis
and optimization by allowing the compiler to control the size of
regions, region formation is unscalable due to aggressive inlining.

\section{Region Formation Analysis with Demand-driven Inlining}
\label{section:demand}

Interprocedural regions that include instructions
from more than one procedure enable
region-based compilation to uncover optimizations 
missed due to procedure boundaries~\cite{Hank97}. 
This section proposes an alternative approach 
to building interprocedural regions which
performs inlining on a demand-driven
basis integrated within region formation analysis
is presented in this section.
By delaying inlining decisions until region formation analysis,
the characteristics of inlined code can be better 
controlled, reducing code growth and memory requirements.
However, inlining performed in this demand-driven way introduces
a number of issues that are not present in existing
region formation techniques; these issues are enumerated, 
and a technique is proposed to addresses them.
In the remainder of this paper, the approach of 
aggressive inlining followed by intraprocedural
profile-sensitive region formation (i.e., Hank, \etal) 
is referred to as \hankreg,
and the new demand-driven approach is called \tomreg.  

\subsection{Challenges in Forming Interprocedural Regions}

Major issues to consider in
the design of \tomreg\ are:

\bheading{Issue 1. Inlining is driven by the \emph{demand} placed at
procedure callsites as regions are formed.}  
Callsites may be encountered as a most frequent successor
or predecessor of a block on a path within the current region
being formed.
The path selection process must determine at that point
whether or not the callee should be inlined,
a decision dependent on the heuristics used
to guide inlining.
If the decision is made to inline a procedure,
it is inlined and region formation
proceeds within the callee's code.
Thus, interprocedural regions are identified by having the
region formation process cross procedure boundaries
by inlining on demand.

\bheading{Issue 2. Region formation analysis must deal with multiple
calls to the same procedure as it crosses procedure boundaries.}
While region formation on the flattened, aggressively inlined code 
of \hankreg\ analyzes a distinct code segment for each callsite 
that has been inlined, 
region formation without prior inlining analyzes the 
same code for a procedure's body for each callsite to that procedure. 
Depending on the context, a callee could be 
partitioned into different regions for different callsites.
\tomreg\ should maintain separate information about a procedure
for each inlinable callsite to that procedure,
or partition the procedure the same each time.

\bheading{Issue 3. The ordering of procedures analyzed 
for region formation and inlining impacts compilation overhead.}
Performing demand-driven inlining can lead to 
large compilation and runtime memory requirements similar to \hankreg\
if the order in which inlining and region formation is performed
is not carefully considered.
As a callsite is encountered in \tomreg, the region formation 
algorithm begins to form regions in the callee. 
Thus, the amount of code growth and the size of data structures needed
during region formation are dependent on
the handling of the worklist of blocks for partitioning 
as region formation crosses procedure boundaries.

\bheading{Issue 4. Procedures may not be inlined at every callsite.}
While a procedure's code is partitioned into regions on demand
at callsites, at some of those callsites the decision may be
made not to inline, resulting in the procedure being partitioned
into local regions in isolation of a calling context. 
Thus, a record of the inlining of each procedure should
be maintained to identify procedures that need to
be processed in isolation during region formation.

\bheading{Issue 5. Total code growth is an imprecise limiting
metric in \tomreg\ since each region will be analyzed and
optimized separately.}
A limit on the memory requirements for \hankreg\ 
is achieved by restricting how large the program 
can grow in total size during the aggressive inlining pass;
however,
individual procedures may be able to grow very large.
This is problematic, since memory requirements during 
analysis are proportional to the size of the largest procedure.
Demand-driven inlining can also 
ensure that individual procedures do not grow excessively 
large by making use of heuristics that consider the impact of 
inlining before it is performed.

\bheading{Issue 6. Region formation may be partially
completed in multiple procedures simultaneously.}
With \tomreg, region formation proceeds recursively.
Region formation starts in a procedure, and when a callsite
is reached it may continue recursively into the callee, 
temporarily suspending region formation in the caller.
Thus, region formation is partially completed in the calling
procedure and will only complete after region formation is
completed in the callee.
When additional levels of recursive region formation occur,
region will be in various stages of
completion along the entire call chain, completing
as each callee invocation returns.

\subsection{A Classification of Regions of a Procedure}

\begin{figure}
\centering
\small
\epsfig{figure=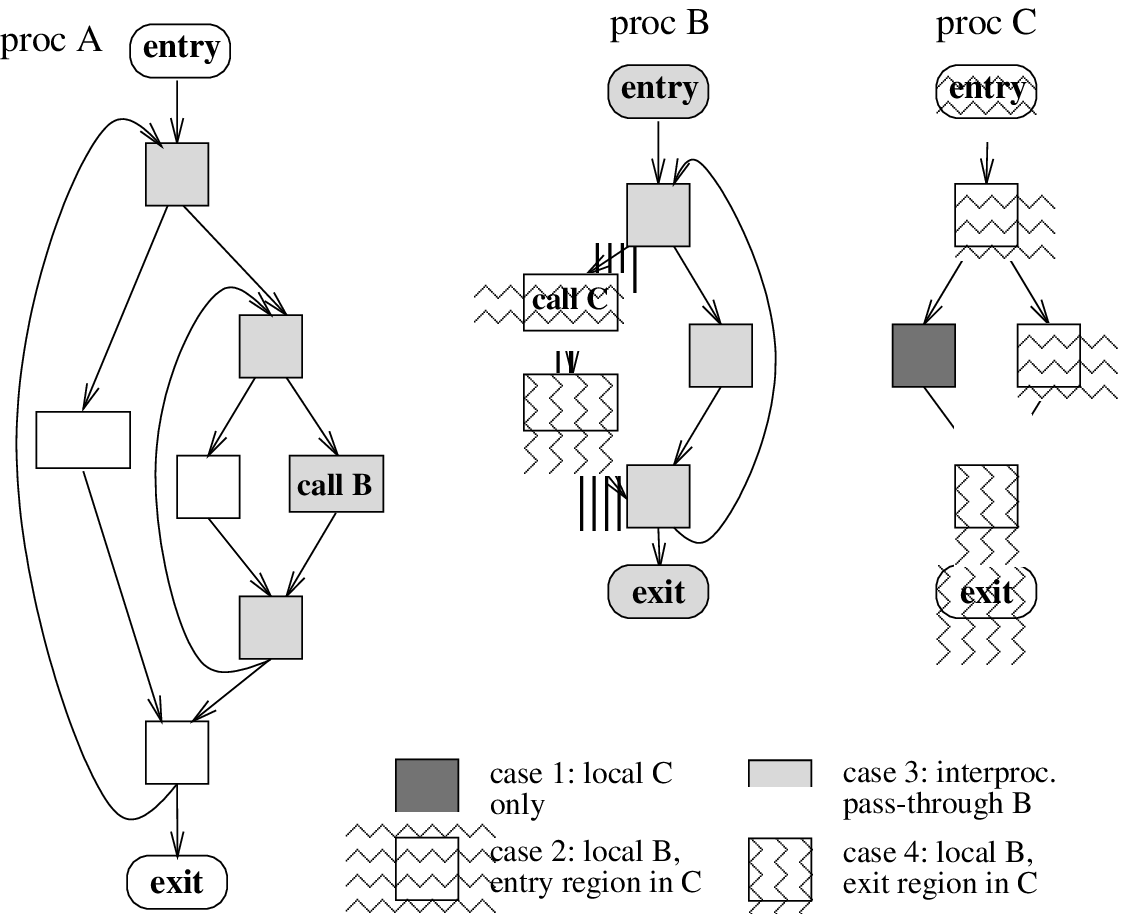,width=3.0in}       
\caption{Illustration of region classification for individual procedures.}
\label{fig:complex}
\end{figure}

The interprocedural region formation algorithm addresses each of the described
issues, based on a classification of regions in a single procedure.  
Regions are classified with respect to individual 
procedures and callsites where they are invoked.
Figure~\ref{fig:complex}, which contains control flow graphs
for three procedures and the formed regions in different
shadings,  illustrates each of the different classifications of regions.
A region in $f$ that includes either the entry or exit block of $f$ is
an interprocedural region.  An interprocedural region can be either entry, exit,
or pass-through.
For each procedure $f$,
each callsite $c$ with a call to 
$f$ has a single \textbf{entry region}
associated with $f$, $entry_{f,c}$
which is the region that contains the entry block of $f$.
At the one callsite in $A$ to procedure $B$ in the figure, the entry region
associated with $B$ contains not only the entry block in $B$ but a path that
passes through to the exit of $B$, and contains the exit of $B$ also.
At the callsite in $B$ to procedure $C$, the entry region associated with $C$
contains the entry block in $C$ and only two other blocks in $C$.

Similarly, each callsite $c$ to procedure $f$ has a single 
\textbf{exit region}, $exit_{f,c}$.
As is the case for the one callsite in $A$ to $B$, 
$entry_{f,c}$ and $exit_{f,c}$ could in fact be the same region because
the region follows a path that passes through from entry to exit; in this
case, it can be said that this region is an interprocedural
\textbf{pass-through region} of $f$ at callsite c.
All remaining regions containing blocks in $f$ 
are \textbf{local regions}, or $local_{f,c}$, 
as they do not involve blocks from the caller of $f$. 
Note that $f$ may not be partitioned into the same regions 
for every callsite to $f$, since region formation within $f$ is based on
the context surrounding the callsite to $f$.

%
%
%
%
%
%

\subsection{An Algorithm for Region Formation with Demand-driven Inlining}

\begin{figure}
\centering
\small
\begin{tabular}{c}
\epsfig{figure=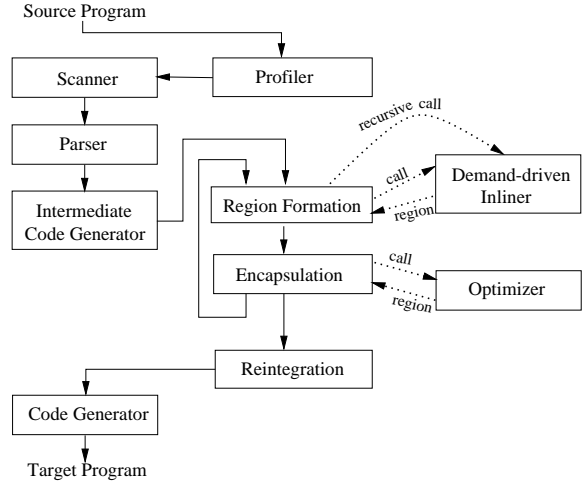,width=3.0in} \\    
\end{tabular}
\caption{Demand-driven region-based compilation framework.}
\label{fig:framework-tom}
\end{figure}

Figure~\ref{fig:framework-tom} presents the organization
of the \tomreg\ framework, and
Figure~\ref{fig:tomsalg} presents the 
region formation algorithm~\cite{Way00}.
\tomreg\ extends \hankreg\ in several important ways in order
to form interprocedural regions without aggressive inlining.
First, when a callsite is encountered as a region is being grown,
{\em FormRegions} recursively calls itself 
to continue to grow the current region in the callee in
the context of the caller, but without inlining at that time. 
Second, in order to minimize the size of the data structures maintained
at any given time during region formation, all regions within a called
procedure will be identified before {\em FormRegions} 
returns to region formation in the caller.  
Third, to enable formation of interprocedural regions through
this recursive approach, {\em FormRegions} operates
on regions rather than just basic blocks.

%
%
\begin{sidewaysfigure}
\small
\input{pact-tomsalgcode}
\vspace{6pt}
\caption{Interprocedural algorithm for region formation with demand-driven inlining}
\label{fig:tomsalg}
\end{sidewaysfigure}

{\em FormRegions} begins with a worklist {\tt B} of all blocks
in the current procedure $f$ for which it is forming regions.
Successor and predecessor blocks are added to the current region 
only if they are desirable as defined in Section \ref{section:background};
{\tt Desirable(x,y)} plays this role.
Non-callsite blocks are appended to the region as  in \hankreg.
When a callsite $c$ is reached in the analyzed code, 
the recursive call to {\em FormRegions} forms regions local 
to the callee, say $g$, and then {\em FormRegions}
returns with the entry and exit regions of $g$. 

If there was \emph{not} a pass-through region of $g$, 
$entry_{g,c}$ is concatenated with the region {\tt R}  
currently being formed in $f$ when the callsite 
was encountered (which completes that interprocedural
region), and this merged region is added
to the local {\tt Rlist} of completed regions in $f$.  
Next, a new region {\tt R} is begun, consisting
solely of $exit_{g,c}$.
If there \emph{is} a pass-through region for $g$, 
this pass-through region is added to {\tt R}, 
but {\tt R} is not necessarily complete at this point. 
Region formation continues in $f$ 
by adding blocks to {\tt R}.
Once all blocks on procedure $f$'s worklist {\tt B}
are exhausted, the return parameters {\tt entryR} 
and {\tt exitR} are assigned the regions in $f$ 
that contain the entry and exit blocks, respectively.
The local regions with respect to $f$ (all regions except the entry and
exit regions of $f$) are optimized and code is generated for them,
prior to returning the entry and exit regions. 

\begin{figure*}
\centering
\small
\begin{tabular}{c@{\hspace{0.35in}}c@{\hspace{0.35in}}c}
\epsfig{figure=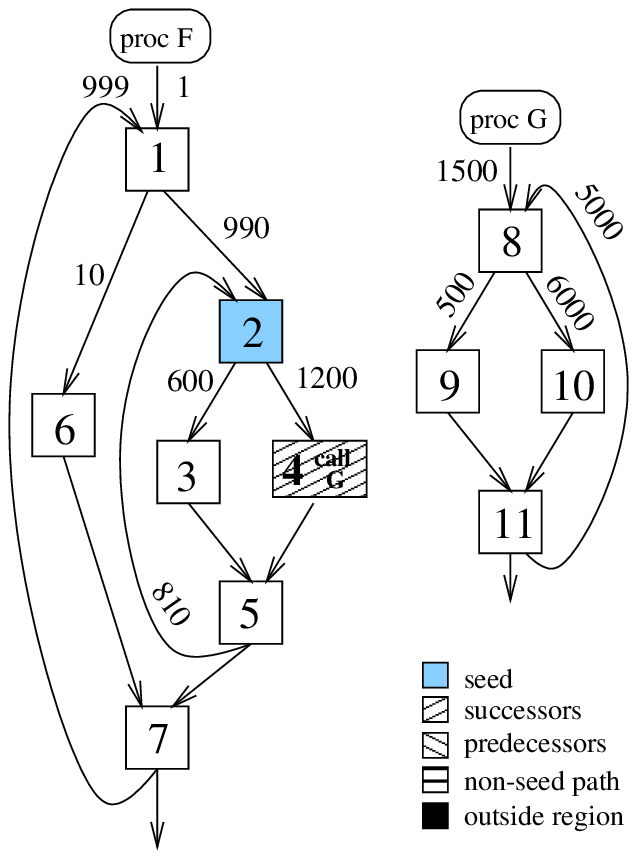,width=1.8in} &
\epsfig{figure=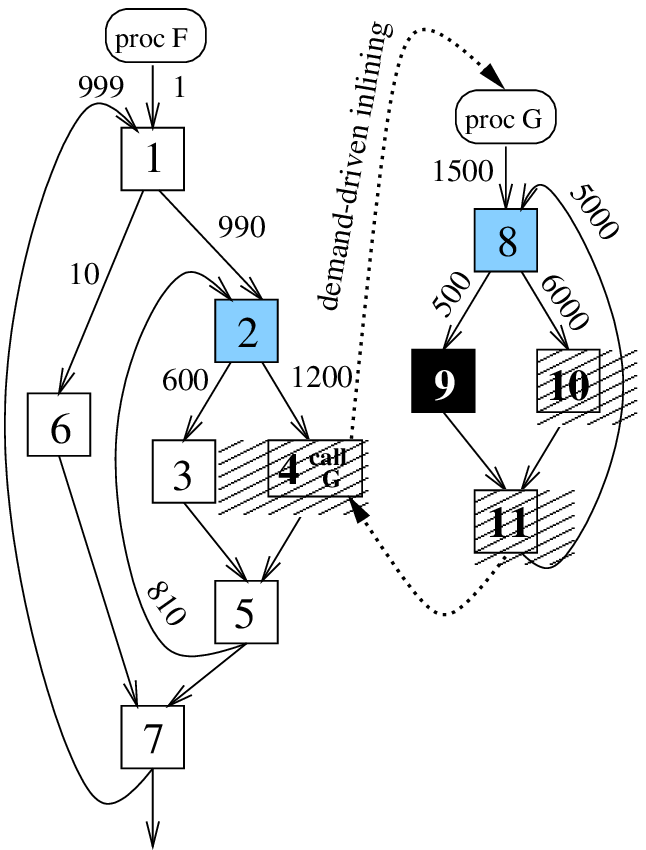,width=1.8in} &
\epsfig{figure=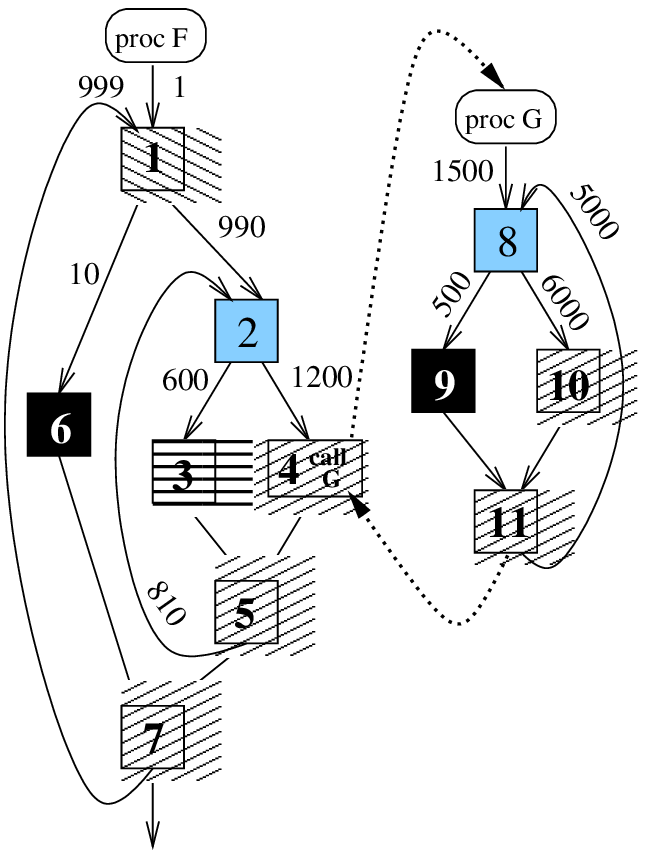,width=1.9in} \\
\multicolumn{1}{p{1.8in}}{\footnotesize\textbf{(a)} 
     Seed (block 2) is selected as it is the
     most frequently executed block in proc F.
     Successors (block 4) are selected until a
     callsite is reached.} &
\multicolumn{1}{p{1.8in}}{\footnotesize\textbf{(b)} 
     Region formation is performed recursively in callee 
     G, where local regions are formed.  Blocks 8, 10 \& 11 
     form one region, with block 9 as a local region.  Region formation 
     then continues in F.} &
\multicolumn{1}{p{1.9in}}{\footnotesize\textbf{(c)} 
     Successor path is completed (5 \& 7), 
     predecessors are added (1), 
     desirable successors are added (3).
     Local regions are formed from remaining blocks (6).
     Region formation is complete.}
\end{tabular}
\caption{Example of \tomreg}
\label{fig:udregion}
\end{figure*}

The main steps of {\em FormRegions} 
are illustrated for a single callsite by the
interprocedural CFGs in Figure~\ref{fig:udregion}.
For clarity, the same fill patterns are used to
differentiate the steps of the \tomreg\ algorithm in this
figure as were used to describe the \hankreg\ algorithm
in Figure~\ref{fig:4steps}.
In this example, a pass-through region of $G$ exists,
is returned to $F$ by {\em FormRegions} as both
{\tt entryR} and {\tt exitR}, and is appended to
the currently forming region {\tt R}.  
Procedures that are not inlined at every callsite, not
inlined at all, or are potential procedure aliases, are
identified after the region formation that began with
the main program is complete. 
The parameter to {\em FormRegions} named {\tt isolated} 
is set for these isolated procedures to indicate
that only local regions are to be formed.

\subsection{Empirical Evaluation}

An experimental comparison of the two
region formation approaches, \tomreg\ and \hankreg,
is described in terms of compilation memory requirements,
code growth and runtime performance.
Analysis of the characteristics of the resulting 
units of compilation, including the size, homogeneity 
of profile weights, and code size
is performed to explain the results.

\subsubsection{Methodology}

These experiments were conducted using the Trimaran compiler 
system~\cite{Trim98}.
With \hankreg\ as an existing component, Trimaran was a 
natural choice for this research.
Significant implementation was performed to add the
capability of demand-driven inlining, 
and to create a region formation module that
incorporates demand-driven inlining and optimization.
Also added was the ability to annotate each basic block
with its procedure of origin to enable identification of
code that was inlined.
For this set of experiments, ten C benchmarks were used
from SPEC 92 and 95 (\texttt{www.spec.org})
representing a variety of computations, 
code sizes and program characteristics.
Table~\ref{tab:3memory} includes numbers of source 
code lines and procedure definitions.

The benchmarks were compiled under three scenarios:
(1)~procedure-based compilation without any 
inlining or region formation, 
(2)~region-based compilation using the \hankreg\ approach, and 
(3)~region-based compilation using the \tomreg\ approach.

\subsubsection{Results}
\label{sec:3results}


\subsubsection*{Compilation memory requirements}

%
%
\begin{table}
\centering 
\small
\caption{Comparison of memory requirements during region formation,
measured in Trimaran \protect\emph{Lcode} instructions.}
\vspace{1ex}
\label{tab:3memory}
\begin{tabular}{|l@{\hspace{6pt}}||r|r||r||r|r|r|r|r|r|} \hline
	& \multicolumn{2}{c||}{\emph{General}}
	& \multicolumn{1}{c||}{\hankreg} 
	& \multicolumn{6}{c|}{\tomreg} \\ \cline{2-10}

	& \multicolumn{2}{c||}{}
	& \multicolumn{1}{c||}{} 
	& \multicolumn{2}{c|}{}
	& \multicolumn{2}{c|}{}
	& \multicolumn{2}{c|}{} \\[-10pt]

	& \multicolumn{1}{c|}{Lines of}
	& \multicolumn{1}{c||}{Num.}
	& \multicolumn{1}{c||}{Memory}
	& \multicolumn{2}{c|}{Static}
	& \multicolumn{2}{c|}{Procedure}
	& \multicolumn{2}{c|}{Memory} \\

	& \multicolumn{1}{c|}{C source}
	& \multicolumn{1}{c||}{of}
	& \multicolumn{1}{c||}{requirement}
	& \multicolumn{2}{c|}{call chain}
	& \multicolumn{2}{c|}{size}
	& \multicolumn{2}{c|}{requirement} \\
Benchmark 
	& \multicolumn{1}{c|}{code}
	& \multicolumn{1}{c||}{procs.}
	& \multicolumn{1}{c||}{Total}
	& \multicolumn{1}{c|}{Avg.}
	& \multicolumn{1}{c|}{Max.}
	& \multicolumn{1}{c|}{Avg.}
	& \multicolumn{1}{c|}{Max.}
	& \multicolumn{1}{c|}{Avg.}
	& \multicolumn{1}{c|}{Worst} \\
\hline \hline
008.espresso  &  14850 &  361 &   73997 &  5 & 11 & 183 & 2059 & 3186 & 5175 \\ \hline
023.eqntott   &   3628 &   62 &   11738 &  3 &  7 & 230 & 1757 & 1156 & 2538 \\ \hline
026.compress  &   1503 &   16 &    2601 &  2 &  5 & 224 & 1761 & 1270 & 1800 \\ \hline
099.go        &  29246 &  383 &  110842 &  9 & 23 & 117 & 1109 & 1076 & 3085 \\ \hline
124.m88ksim   &  19092 &  252 &   55783 &  6 & 11 & 193 & 1537 & 1195 & 1923 \\ \hline
126.gcc       & 205627 & 1170 & 1050754 &  5 & 13 & 202 & 1810 & 2666 & 4391 \\ \hline
130.li        &   7597 &  357 &   31552 & 22 & 35 & 112 &  987 & 1640 & 3197 \\ \hline
132.ijpeg     &  29259 &  473 &  112188 &  8 & 14 & 124 & 2510 & 1385 & 2185 \\ \hline
134.perl      &  27044 &  316 &  100063 &  5 & 15 & 140 & 1977 & 1498 & 2732 \\ \hline
147.vortex    &  67202 & 1127 &  302409 &  4 & 12 & 131 & 2301 & 1166 & 2210 \\ \hline
\hline
\textbf{average} & 40505 & 452 &  17363 &  6 & 11 & 162 & 1397 & 1274 & 2228 \\ \hline
\end{tabular}
\end{table}

Table~\ref{tab:3memory} compares the compilation memory
requirements for \hankreg\ versus \tomreg.
Due to design considerations of the Trimaran framework,
direct measurement of memory requirements was not possible.  
Instead, measurements of whole program size, procedure sizes,
and static call chain lengths were taken, and estimates
of memory requirements were computed according to each
strategy for region-based compilation.

For \hankreg, the compilation memory requirements are
computed as code size after aggressive inlining
is performed, as measured in number of \emph{Lcode} instructions,
because the entire program may be held in memory 
during region formation and optimization (in the worst case).
For \tomreg, first the average and maximum sizes
of procedures in a benchmark were calculated.
Next, the lengths of static acyclic call chains were 
measured at the source code level.
The call chain length and procedure size information
were then used to compute the average and maximum
of the sum of procedure sizes along the average and
maximum length call chains.
The average value provides a good estimate of typical 
compilation memory usage for purposes of comparison, 
while the maximum value indicates the worst case.

\begin{table}
\centering 
\small
\caption{Percentage difference in average and maximum
memory requirements of \hankreg\ and \tomreg.}
\vspace{1ex}
\label{tab:3avgmemory}
\begin{tabular}{|l@{\hspace{6pt}}||r|r|} \hline
	& \multicolumn{2}{c|}{} \\[-10pt]
	& \multicolumn{2}{c|}{$\frac{\tomreg}{\hankreg}$~\%} \\
Benchmark	& average & maximum \\
\hline \hline
008.espresso  &  4.3 &  7.0 \\ \hline
023.eqntott   &  9.8 & 21.6 \\ \hline
026.compress  & 48.8 & 69.2 \\ \hline
099.go        &  1.0 &  2.8 \\ \hline
124.m88ksim   &  2.1 &  3.4 \\ \hline
126.gcc       &  0.3 &  0.4 \\ \hline
130.li        &  5.2 & 10.1 \\ \hline
132.ijpeg     &  1.2 &  1.9 \\ \hline
134.perl      &  1.5 &  2.7 \\ \hline
147.vortex    &  0.4 &  0.7 \\ \hline
\hline
\textbf{average} & 7.5 & 12.0 \\ \hline
\end{tabular}
\end{table}

The data in Table~\ref{tab:3avgmemory}
shows that on average, \tomreg\ uses about 7.5\% of the 
memory required by \hankreg\ for region formation for the 
benchmarks studied, over a range of roughly $<$1\% to 49\%.
In the worst case, \tomreg\ uses an average of 12\% of
the memory required by \hankreg\ over a range of
about $<$1\% to 69\%.
Benchmarks with larger numbers of procedures
and procedure calls, and more and longer call chains,
benefited the most from \tomreg.  While smaller benchmarks
showed some benefit, the smallest,
\emph{026.compress}, showed the least benefit, suggesting
that \tomreg\ may be best suited to large applications.

%

\subsubsection*{Code growth}

Code growth was measured as the percentage change in 
overall code size from the original program,
shown in Table~\ref{tab:3codegrowth} as the percentage
increase or decrease in size.
To measure their code size used to calculate code growth, 
each benchmark was compiled in three ways:
(1) without any inlining or region formation, 
(2) using the \hankreg\ strategy, and 
(3) using the \tomreg\ strategy.
Measurements were taken in terms of \emph{Lcode} instructions
of the resulting compiled programs.
An increase in code size is represented by a positive value.
For example, a value of 21 means that after
compilation within a particular framework,
the program is 21\% larger than the same program
compiled using the procedure-based approach.

%
%
\begin{table}
\centering
\small
\caption{Percentage change in code growth for \hankreg\ and \tomreg.}
\vspace{1ex}
\label{tab:3codegrowth}
\begin{tabular}{|l@{\hspace{6pt}}||r|r||r|} \hline
	& 
	&
	& \multicolumn{1}{c|}{\tomreg} \\
Benchmark 
	& \multicolumn{1}{c|}{\hankreg}
	& \multicolumn{1}{c||}{\tomreg}
	& \multicolumn{1}{c|}{$-$~\hankreg} \\
\hline \hline
008.espresso & 21 & 19 & -2  \\ \hline
023.eqntott  & 24 & 26 & +2  \\ \hline
026.compress & 26 & 25 & -1  \\ \hline
099.go       & 22 & 25 & +3  \\ \hline
124.m88ksim  & 21 & 20 & -1  \\ \hline
126.gcc      & 22 & 23 & +1  \\ \hline
130.li       & 20 & 23 & +3  \\ \hline
132.ijpeg    & 21 & 24 & +3  \\ \hline
134.perl     & 22 & 23 & +1  \\ \hline
147.vortex   & 21 & 21 & 0  \\ \hline
\textbf{average} & 22.0 & 22.9.1 & +0.9 \\ \hline
\end{tabular}
\end{table}

On average, \tomreg\ introduces $<1$\% more code than \hankreg,
over a range of 2\% less to 3\% more growth.
In general, differences in code growth are not dramatic,
due to the use of the global static code growth limit of 20\%
in both \hankreg\ and \tomreg.
In practice, the 20\% code growth limit prevents
inlining once the code size has grown to 20\% or
more above the original size.  
However, a benchmark may grow to just below this limit,
allowing one more instance of inlining to be performed.  
\tomreg\ shows slightly more code growth than \hankreg\
because \tomreg\ is inlining in a different order, which can lead to
the benchmark first growing to just below the limit,
and then inlining a larger procedure which exceeds the
limit considerably.


\subsubsection*{Runtime performance}

Table~\ref{tab:3compruntime} reports the percentage change in
execution time.
Negative values for percentage change in 
execution time indicate a performance speedup; the
program ran faster compared to the procedure-based compilation.
The last column shows the difference
in the change in execution time between \tomreg\ and \hankreg,
with a negative value indicating that a benchmark compiled
using \tomreg\ ran faster than when compiled with \hankreg;
a positive difference indicates that \hankreg\ was faster.

%
%
\begin{table}
\centering
\small
\caption{Percentage change in execution time
for \hankreg\ and \tomreg\ compared to procedure-based.}
\label{tab:3compruntime}
\begin{tabular}{|l@{\hspace{6pt}}|r|r||r|} \hline
	& 
	& 
	& \multicolumn{1}{c|}{\tomreg} \\
Benchmark
	& \multicolumn{1}{c|}{\hankreg}
	& \multicolumn{1}{c||}{\tomreg}
	& \multicolumn{1}{c|}{$-$~\hankreg} \\
\hline \hline
008.espresso &  -6.13 & -1.12 &  5.01 \\ \hline
023.eqntott  &  -3.17 & -2.14 &  1.03 \\ \hline
026.compress &  -3.11 & 26.88 & 29.99 \\ \hline
099.go       &  -6.28 &  7.30 & 13.58 \\ \hline
124.m88ksim  &  -4.65 & -2.40 &  2.25 \\ \hline
126.gcc      &  -6.72 & -5.00 &  1.72 \\ \hline
130.li       &  -8.49 & 12.50 & 20.99 \\ \hline
132.ijpeg    &  -7.01 & -5.99 &  1.02 \\ \hline
134.perl     &  -4.22 & -2.18 &  2.04 \\ \hline
147.vortex   &  -6.90 & -3.72 &  3.18 \\ \hline
\hline
\textbf{average} & -5.67 & 2.41 &  8.08 \\ \hline
\end{tabular}
\end{table}

For seven of the ten benchmarks, 
the results for execution time were quite similar 
for \hankreg\ and \tomreg, separated only by a few
percentage points, which equates to fractions of a second
in wall clock time.  
In particular, there are little or no differences in performance for 
\emph{008.espresso}, \emph{023.eqntott}, \emph{124.m88ksim},
\emph{126.gcc}, \emph{132.ijpeg}, \emph{134.perl} and \emph{147.vortex}.
The drop in performance from \hankreg\ to \tomreg\ for
\emph{026.compress}, \emph{099.go} and \emph{130.li} 
is due to naive heuristics for deciding whether to
perform demand-driven inlining at a given callsite, 
and the way the prototype system handles demand-driven 
inlining of indirect recursive procedure calls. 
Specifically, with this implementation, it is possible for
the code limit to be reached before inlining is performed in some of the 
high execution frequency regions, resulting in optimization loss. 
ILP processor utilization was also examined, with only
insignificant variations noted.

Thus, while memory requirements are improved dramatically,
runtime performance remains virtually unaffected in general.
This improvement in memory requirements was the 
primary goal of performing demand-driven inlining
during region formation in \tomreg.
Since \tomreg\ is implemented using the same region
formation and inlining heuristics, leading to substantially
similar regions, dramatic improvements to runtime
performance could not be reasonably expected.
The key innovation of \tomreg\ is to integrate
demand-driven inlining into region formation
to reduce the requirements for memory during compilation.

\subsection{Analysis of Compilation Unit Characteristics}

Procedure restructuring affects the characteristics
of the unit of compilation.
Analyzing changes to program characteristics, such
as the size, profile homogeneity and interprocedural
scope of the unit of compilation,
can further explain the impact on memory requirements,
code growth and performance.

\subsubsection*{Unit size}

%
%
\begin{table}
\centering
\small
\caption{Comparison of number of 
compilation units for procedure-based, \hankreg\ and \tomreg\
(in \protect\emph{Lcode} instructions).}
\label{tab:3numunits}
\begin{tabular}{|l@{\hspace{6pt}}||r|r|r||r|} \hline
&
	\multicolumn{3}{c}{Number of Units} &
	\tomreg \\
Benchmark &
	{\footnotesize Proc.-based}&
	{\footnotesize \hankreg} &
	{\footnotesize \tomreg} &
	$-$~\hankreg \\
\hline \hline
008.espresso &  361 & 1787 & 1774 & -13 \\ \hline
023.eqntott  &   62 &  436 &  476 &  40 \\ \hline
026.compress &   16 &  117 &  102 & -15 \\ \hline
099.go       &  383 & 1838 & 1888 &  50 \\ \hline
124.m88ksim  &  252 & 1336 & 1322 & -13 \\ \hline
126.gcc      & 1170 & 6084 & 6047 & -37 \\ \hline
130.li       &  357 &  801 &  793 &  -8 \\ \hline
132.ijpeg    &  473 & 3575 & 3791 & 216 \\ \hline
134.perl     &  316 &  822 &  797 & -25 \\ \hline
147.vortex   & 1127 & 5522 & 5616 & 94 \\ \hline
\hline
\textbf{average} & 452 & 2232 & 2261 & 29 \\ \hline
\end{tabular}
\end{table}

Tables~\ref{tab:3numunits} and \ref{tab:3unitsize} 
report the total number of compilation units and
average size in \emph{Lcode} instructions
for each of the studied benchmarks
under the three different strategies for compilation.  
The two region-based compilation techniques result in very similar
average region size and total number of regions,
while the procedure-based strategy produces far fewer, though
far larger, compilation units.
Slight variations in sizes and numbers of regions
are attributed to differences in the order in which 
callsites are inlined.  
The aggressive inlining
of \hankreg\ favors inlining frequently executed,
smaller procedures over larger procedures due to
the limit it places on total code growth and
the inlining heuristic.
Since the demand-driven inliner inlines as it is 
creating a region and reaches a callsite, it can reach the
same specified limit for code growth at a different time
due to different order of inlining.
The demand-driven approach to inlining in \tomreg\ 
and the recursive nature of the algorithm lead
to a bottom-up inlining of regions.
That is, the inlining is performed as the recursive
calls to \texttt{FormRegions} return.
The contribution of the \tomreg\ approach is that 
it can significantly reduce compilation memory requirement,
while creating number and size of regions comparable to
those created by \hankreg.

%
%
\begin{table}
\centering
\small
\caption{Comparison of average size of 
compilation units for procedure-based, \hankreg\ and \tomreg\
in \emph{Lcode} instructions).}
\label{tab:3unitsize}
\begin{tabular}{|l@{\hspace{6pt}}||r|r|r||r|} \hline
&
	\multicolumn{3}{c}{Average unit size} &
	\tomreg \\
Benchmark &
	{\footnotesize Proc.-based} &
	{\footnotesize \hankreg} &
	{\footnotesize \tomreg} &
	$-$~\hankreg \\
\hline \hline
008.espresso & 183 & 50 & 49 & -1 \\ \hline
023.eqntott  & 169 & 33 & 31 & -2 \\ \hline
026.compress & 152 & 28 & 32 & +4 \\ \hline
099.go       & 234 & 55 & 51 & -4 \\ \hline
124.m88ksim  & 155 & 31 & 30 & -1 \\ \hline
126.gcc      & 530 & 62 & 57 & -5 \\ \hline
130.li       &  81 & 47 & 49 & +2 \\ \hline
132.ijpeg    & 137 & 38 & 37 & -1 \\ \hline
134.perl     & 206 & 51 & 49 & -2 \\ \hline
147.vortex   & 161 & 35 & 34 & -1 \\ \hline
\hline
\textbf{average} & 201 & 43 & 42 & -1 \\ \hline
\end{tabular}
\end{table}

\subsubsection*{Profile homogeneity}

\emph{Profile homogeneity}
is defined as the measure of how similar the given
unit of compilation is in terms of profile weight
per instruction, operation or basic block.
This variation on code density provides an indicator for
the impact of region formation on optimization.  
More homogeneous compilation units enable the optimizer
to easily identify and isolate heavily executed regions,
and then selectively focus more attention on these
more important regions and less attention elsewhere.
This partitioning reduces the chance of
leaving important portions of the
code unoptimized or spending excessive time optimizing
unimportant code.

Within the context of units of compilation,
the profile homogeneity, or \emph{profile variance}, 
is defined to be the measure of the degree of
deviation, that is, the standard deviation,
in profile weights for all basic
blocks within a compilation unit.
Table~\ref{tab:3homogen} shows the average profile variance
and percentage of compilation units that
are invariant for each benchmark.
The \emph{average profile variance} is an overall indication of how
consistent the profile weights are within each
of the benchmarks' units of compilation.
The closer the profile variance is to 0, the
less variation there is in the profile weights 
overall for the benchmark, and the more 
homogeneous the benchmark.

%
%
\begin{table}
\centering
\small
\caption{Comparison of percentage of invariant compilation units 
and profile variance (homogeneity) for 
procedure-based, \hankreg\ and \tomreg.}
\label{tab:3homogen}
\footnotesize
\begin{tabular}{|l@{\hspace{6pt}}||r|r||r|r||r|r|} \hline
 & 
	\multicolumn{2}{c||}{Proc.-based} & 
	\multicolumn{2}{c||}{\hankreg} & 
	\multicolumn{2}{c|}{\tomreg} \\
&
	\multicolumn{1}{c|}{Profile} & \multicolumn{1}{c||}{Pct. units} &
	\multicolumn{1}{c|}{Profile} & \multicolumn{1}{c||}{Pct. units} &
	\multicolumn{1}{c|}{Profile} & \multicolumn{1}{c|}{Pct. units} \\
Benchmark &
	\multicolumn{1}{c|}{variance} & \multicolumn{1}{c||}{invariant} &
	\multicolumn{1}{c|}{variance} & \multicolumn{1}{c||}{invariant} &
	\multicolumn{1}{c|}{variance} & \multicolumn{1}{c|}{invariant} \\
\hline \hline
008.espresso & 0.362 & 88.7 & 0.340 & 81.2 & 0.342 & 94.1 \\ \hline
023.eqntott  & 0.017 & 96.1 & 0.001 & 97.6 & 0.020 & 97.6 \\ \hline
026.compress & 0.313 & 90.7 & 0.245 & 87.5 & 0.375 & 90.8 \\ \hline
099.go       & 0.292 & 87.1 & 0.293 & 88.2 & 0.293 & 91.0 \\ \hline
124.m88ksim  & 0.272 & 92.1 & 0.249 & 92.9 & 0.292 & 93.3 \\ \hline
126.gcc      & 0.132 & 89.0 & 0.108 & 90.1 & 0.108 & 93.2 \\ \hline
130.li       & 0.198 & 90.3 & 0.208 & 88.5 & 0.203 & 94.2 \\ \hline
132.ijpeg    & 0.273 & 88.8 & 0.254 & 86.7 & 0.310 & 91.1 \\ \hline
134.perl     & 0.212 & 88.7 & 0.195 & 89.3 & 0.187 & 90.3 \\ \hline
147.vortex   & 0.310 & 90.7 & 0.259 & 91.1 & 0.261 & 93.1 \\ \hline
\hline
\hline
\textbf{average} & 0.238 & 90.2 & 0.215 & 89.3 & 0.239 & 92.9 \\ \hline
\end{tabular}
\end{table}

%

The results in Table~\ref{tab:3homogen} indicate
that in every case \tomreg\ improves percentage of invariant
units over both procedure-based compilation and \hankreg.
\hankreg\ tended to gain in some cases and lose in others
over procedure-based compilation.
When there is an increase in the
percentage of invariant code, there is generally
also an increase in the profile variance of the code overall.
This is due to the procedure restructuring done by region
formation, which favors grouping more frequently executed
code together, leaving less frequently executed code
behind.  
Because less important code is not 
actively formed into more homogeneous regions,
the profile weights of their containing regions
are slightly more variant than regions of
frequently executed code.
It can be hypothesized that
\tomreg\ produces less variant code over \hankreg\ because 
the integrated, demand-driven
use of inlining within \tomreg\ uses
the region formation \emph{desirability} heuristic
(50\% or greater execution frequency) to also
guide inlining.  Overall, the frequency of code inlined
by \tomreg\ is likely to be greater than the
more general, aggressive inlining approach in \hankreg.

\subsubsection*{Interprocedural scope}

When specifically comparing regions to procedures,
and regions formed using different techniques,
a change in the number of \emph{interprocedural regions}
and the amount of \emph{interprocedural operations} per
region indicates the change in interprocedural scope.
Recall that an interprocedural region is a region that
includes instructions from more than one procedure.
Interprocedural operations are the instructions in
an interprocedural region that are from procedures other
than the procedure in which formation of the region began.
Before inlining is performed, all basic blocks 
are annotated with the block's procedure of origin.
With this origin information, the impact of region formation
on interprocedural scope in a unit of compilation can be 
measured directly by calculating how much of the code 
within each unit originated outside itself.
The percentage of interprocedural code in the program is
measured as a simple ratio of the number
of interprocedural operations to total operations.

Table~\ref{tab:3interscope} shows the average percent
of code within regions that is from a procedure outside the
region (i.e., interprocedural code).
An improvement in
the percentage is indicative of better interprocedural scope.
An increase in interprocedural scope within the
unit of compilation means that the potential for
interprocedural optimization is increased without additional analysis.

%
%
\begin{table}
\centering
\small
\caption{Comparison of percentage of interprocedural 
operations in \hankreg\ and \tomreg.}
\label{tab:3interscope}
\begin{tabular}{|l@{\hspace{6pt}}||r|r||r|} \hline
& 
	\multicolumn{1}{c|}{\footnotesize \hankreg} & 
	\multicolumn{1}{c||}{\footnotesize \tomreg} &
	\tomreg \\
Benchmark & \multicolumn{1}{c|}{\footnotesize \% interproc.} & 
	\multicolumn{1}{c||}{\footnotesize \% interproc} &
	$-$~\hankreg \\
\hline \hline
008.espresso & 20.7 & 24.3 &  3.6 \\ \hline
023.eqntott  & 18.0 & 23.8 &  5.8 \\ \hline
026.compress & 23.5 & 25.9 &  2.4 \\ \hline
099.go       & 28.4 & 26.7 & -1.7 \\ \hline
124.m88ksim  & 22.9 & 24.9 &  2.0 \\ \hline
126.gcc      & 19.3 & 25.4 &  6.1 \\ \hline
130.li       & 30.2 & 28.0 & -2.2 \\ \hline
132.ijpeg    & 23.0 & 25.1 &  2.1 \\ \hline
134.perl     & 20.0 & 24.9 &  4.9 \\ \hline
147.vortex   & 23.0 & 25.7 &  2.7 \\ \hline
\hline
\textbf{average} & 22.9 & 25.5 & 2.6 \\ \hline
\end{tabular}
\end{table}

In general, the percentage of interprocedural operations
is similar for \hankreg\ and \tomreg.
The differences in interprocedural scope under the
two techniques are slight.
The interaction of various factors leads to insight on how
to improve the techniques.
The slight increase in interprocedural scope for
\emph{008.espresso} occurs 
with a slight decrease in code growth and little
or no change to profile variance.
For \emph{023.eqntott}, slight differences in 
code growth and variance would not indicate
the larger increase in interprocedural scope seen
for \tomreg.  This change could be due to the
slight reduction seen for average size of the
unit of compilation for \tomreg\ for \emph{023.eqntott},
since the other factors were quite similar.
The improvements to interprocedural scope seen with 
\emph{126.gcc}, \emph{132.ijpeg}, \emph{134.perl} and \emph{147.vortex}
are likely due to slight decreases in the average
size of the unit of compilation, which are magnified
due to the large sizes of the benchmarks.
Most puzzling is the increase in interprocedural scope
seen with \tomreg\ applied to \emph{026.compress},
\emph{099.go}, \emph{130.li}, and to a lesser extent, \emph{124.m88ksim}, 
which exhibit significantly more variance and
a definite reduction in runtime performance versus
\hankreg.  This seeming contradiction for these four benchmarks
could be due to the effect of gaining interprocedural scope by
restructuring, with the side-effect of leaving
behind more invariant code in the process.
An increase in code growth appears to be the cause of
decreased interprocedural scope for \emph{130.li}.

\section{Heuristics for Demand-driven Inlining}
\label{section:heuristics}

In the previous section, baseline heuristics were used to guide
demand-driving inlining within region formation to establish the
efficacy of the \tomreg\ approach as compared with \hankreg\ and
traditional procedure-based compilation.
This section explores a variety of heuristics designed
to improve the performance of \tomreg, and discusses a number of classifications,
factors and important issues that are integral to inlining heuristics design.

\begin{figure}
\centering
\small
\epsfig{figure=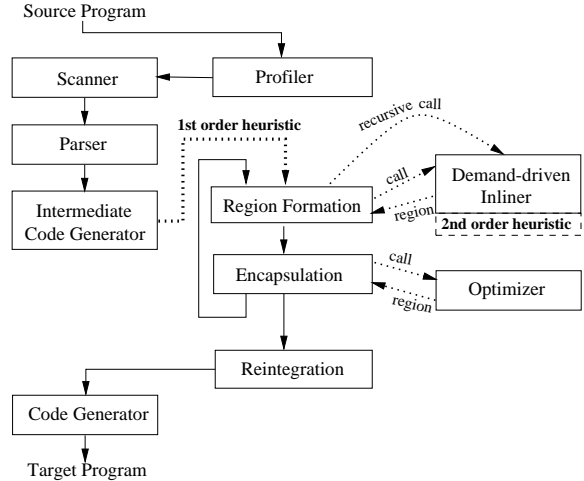,width=3.0in}   
\caption{Relationship of first- and second-order heuristics to
region formation and the demand-driven inliner.}
\label{fig:twoheuristic}
\end{figure}

Because region formation drives inlining, the heuristics
for a demand-driven inliner must consider 
the order that procedures are processed by the region
formation phase and the characteristics of a callee
at each callsite as it is encountered during region formation~\cite{Way01a}.
Procedures that are analyzed later in the compilation may 
result in less inlined code within them and thus be less optimized
since code growth restrictions could limit further inlining,
and thus limit the interprocedural scope of that procedure.
Therefore, procedures that have the highest potential for optimization,
particularly instruction scheduling for ILP architectures,
should be processed first by the region formation analysis phase.
Thus, demand-driven inlining within a region-based compiler 
involves two general classes of heuristics, defined as:
\emph{first-order} heuristics that
determine the order in which procedures are processed during
region formation, 
and \emph{second-order} heuristics that govern
decisions about whether to inline 
at each callsite.
Figure~\ref{fig:twoheuristic} illustrates the location
within a region-based compilation framework of these two heuristics.

\subsection{First-order Heuristics}

\emph{First-order heuristics} 
select the order to consider procedures for region formation, 
which will implicitly affect the order of 
demand-driven inlining decisions.  
Because demand-driven inlining within a given procedure
is considered at callsites as region formation is performed 
for that procedure,
the order of decisions for demand-driven inlining follows
the flow-directed manner in which region formation
is performed within a given procedure's control flow graph.


The first-order heuristics studied in the research
attempt to order procedures from most to least
important in terms of optimization opportunity.  
In particular, three possible first-order
heuristics for demand-driven inlining were examined.
The most precise measurement of
procedure importance is actual dynamic run-time
profiling which comes at the cost
of an initial instrumentation, compilation and
execution.  
Procedures are ordered from highest to lowest 
percentage of overall run-time spent in the procedure, based
on profiling information.  
It is worth noting that procedures which consume
larger portions of execution time are likely to
contain loops and callsites within the loops,
which supports the importance of this heuristic
to interprocedural region formation in a demand-driven framework.

Static estimates of importance provide
less costly heuristics, but also tend
to produce less precise information.  
One heuristic based on static estimates
orders procedures from most to least number of 
static callsites within the procedure,
and within that order from smallest to largest procedure size.
More importance is assigned to 
procedures with the highest percentage of callsites
compared with code size.  This increases the chance
that region formation will be performed interprocedurally,
producing more scheduling and optimization opportunities,
while controlling code growth by considering
smaller procedures before larger ones.

Another ordering considered 
is based on the {\em loop call weight} of a procedure,
assigning more
importance to procedures which contain more callsites
within loops, and increased importance for those
callsites that are more deeply nested. 
The loop call weight is computed as:
$\sum_{i=1}^{n} loopdepth_{i} \times W$, where
$n$ is the number of callsites in a procedure,
and $W$ is the loop depth weight constant.
A value of 10 is used for $W$ to assign an order 
of magnitude increase in significance
to successive loop depths, since, intuitively, 
interior loops consume more execution cycles than 
do their enclosing loops.

\subsection{Second-order Heuristics}

\emph{Second-order heuristics} involve the decision about
whether to inline each callee within a procedure
as region formation reaches that callsite.
While there are a number of heuristics already developed
for this decision making, they have all been applied
within a separate inlining phase without consideration
of the interactions with region formation analysis,
and in particular, demand-driven inlining.
The second-order heuristics attempt
to increase instruction scheduling and optimization
opportunities 
while minimizing code growth.
For correctness, procedures
where there are mismatches in the
number and types of parameters between the callsite
and callee, when the compiler determines that 
memory regions associated with arguments to a procedure may overlap
or are pointers, are not inlined.

To avoid high code growth, inlining is prevented once the 
the overall code size has increased more than
20\% percent above the original size.
A code growth limit of 20\%
has been shown to minimize unnecessary code growth while still allowing
beneficial inlining~\cite{Hank97}.
Similarly, inlining is prevented for procedures that are directly or
indirectly recursive to avoid
the potential for excessive code growth.

Procedures that are more frequently
executed than a fixed frequency or with some desired ratio over
the frequency of the caller are inlined. 
Region formation already uses this 
second-order heuristic, such that inlined procedures
will always be executed at least
50\% as frequently as the seed block of their
enclosing region.
Only procedures that are less than a static
maximum size are inlined to limit code growth,
and procedures with higher call overhead compared
with their code size are inlined.

\subsection{Empirical Evaluation}

Experiments were conducted to study the effectiveness 
of a number of heuristic combinations (Table~\ref{tab:heuristics}),
and to determine which
strategies can improve characteristics of the program
and its runtime performance.
The heuristic combinations were compared
by measuring three effects in terms of the  
percentage change of each combination versus \Hzero,
the baseline method.
In particular, the effects that were evaluated include: 
(1) \emph{code growth}, 
(2) \emph{compilation time}, 
including the time to compile the source code
up through region formation and region-based optimization,
and (3) \emph{execution time}, which measures
more directly the impact of inlining heuristics
on region formation and region-based optimization, 
and ultimately on runtime performance.
Note that the results in this section cannot be compared
directly to those in Section~\ref{section:demand},
due to slight variations in implementation needed to
incorporate the newer inlining heuristics.

\begin{table}
\centering
\small
\caption{Summary of heuristic combinations.\label{tab:heuristics}}
\vspace{1ex}
\footnotesize
\begin{tabular}{|l||p{0.2\textwidth}|p{0.2\textwidth}|p{0.4\textwidth}|} \hline
Name
	& \multicolumn{1}{c|}{First-order} 
	& \multicolumn{1}{c|}{Second-order}
	& \multicolumn{1}{c|}{Intuition/Motivation} \\ 
\hline \hline
\Hzero 
	& None
	& None
	& Baseline version of original region-based compilation.
	No inlining is performed. \\
\hline
\Hone 
	& Run-time profile ordering.
	& Inlined procedures guaranteed to be 
	executed at least 50\% as frequently as seed block in 
	their region~\cite{Hank97}.
	& Original region-based compilation.
	Aggressive inlining with standard code growth limit, then
	region formation; first- and second-order inlining heuristics 
	as defined by~\cite{Hank97}. \\
\hline
\Htwo 
	& Order by descending number of static 
	callsites, then ascending procedure size.
	& Same as \Hone, plus only inline if 
	callee size $\leq$~25~\cite{Hall95}.
	& Demand-driven inlining with simple static heuristics;
	avoid more costly analysis in order to potentially improve
	compilation time.  \\
\hline 
\Hthree 
	& Same as \Htwo.
	& Same as \Hone, plus prevent inlining 
	direct or indirect recursion. 
	& Demand-driven inlining with simple static heuristics.
	Increase number of procedures into which inlining is performed
	before code growth limit is reached, preventing
	successive inlining of recursive procedures. \\
\hline
\Hfour 
	& Order by decreasing loop call weight, 
	then ascending procedure size.
	& Same as \Hthree.
	& Static estimation of profile information by equating
	loop characteristics with predicted execution frequency,
	for improved compilation time. \\
\hline
\Hfive 
	& Order by decreasing execution cycles, 
	then ascending procedure size.
	& Same as \Hthree.
	& Actual runtime profile information should provide most precise
	information for guiding region formation and demand-driven
	inlining, for improved runtime performance. \\ 
\hline
\Hsix 
	& Same as \Hfive.
	& Same as \Hthree, plus minimum loop call weight of 10 to inline. 
	(Note: a procedure containing a single loop is assigned
	a loop call weight of 10.)
	& Only inline if contains at least 1 call within at least one loop.
	Combines profile information to prioritize compilation of
	procedures, with potentially faster static loop characteristic 
	estimation for making inlining decisions; should improve
	compilation time. \\ 
\hline
\end{tabular}
\normalsize
\end{table}

When designing inlining heuristics,
first-order heuristics should not ignore the goals of
second-order heuristics.
In particular, first-order heuristics should anticipate 
second-order heuristics by processing procedures early in the
compilation that will benefit most from the interprocedural scope
gained from demand-driven inlining.
Second-order heuristics should rely on first-order
heuristics to provide more important procedures
earlier in the compilation, while constraining
code growth so that procedures remaining to be
handled by region formation can still benefit
from demand-driven inlining.
While the heuristics are the same for \hankreg\ in 
Section~\ref{section:demand} and \Hone\ in this section, for example,
the implementation of the heuristics was modified 
to enable consistent comparison of results with the newer heuristics.
Experimental results reported in Section~\ref{section:demand} enable
the initial valid comparison of \tomreg\ with the original,
unmodified \hankreg\ framework.

\subsubsection{Methodology}

Implementation of the described techniques and experiments 
has been conducted in context of the Trimaran compiler~\cite{Trim98}.
The existing region formation module was enhanced to 
incorporate additional first-order inlining heuristics.
The demand-driven inliner within \tomreg\ was
extended with a number of new second-order inlining heuristics.
In addition, the demand-driven inliner was more tightly
integrated into the compiler, enabling a meaningful
measurement of compilation time.
The experiments were performed on the same set of 
benchmarks (Table~\ref{tab:3memory}, p.~\pageref{tab:3memory}).

\subsubsection{Results}

\subsubsection*{Code growth}

Table~\ref{tab:heuristics-codegrowth} reports the percentage
increase in code growth for heuristics \Hone\ through \Hsix\
versus the baseline compilation \Hzero.
Code growth was measured directly by counting the
number of \emph{Lcode} instructions resulting from
compilation using each of the heuristics.

\begin{table}
\centering
\small
\caption{Percentage change in code growth over \Hzero.}
\vspace{1ex}
\label{tab:heuristics-codegrowth}
\begin{tabular}{|l@{\hspace{6pt}}|r|r|r|r|r|r|} \hline
Benchmark & \Hone & \Htwo & \Hthree & \Hfour & \Hfive & \Hsix \\
\hline \hline
008.espresso &  8 & 1 & 20 & 20 & 17 & 17 \\ \hline
023.eqntott  &  6 & 0 & 18 & 18 & 15 & 15 \\ \hline
026.compress & 17 & 0 & 23 & 23 & 21 & 21 \\ \hline
099.go       &  9 & 3 & 12 & 11 &  8 &  8 \\ \hline
124.m88ksim  &  8 & 1 & 16 & 15 & 12 & 12 \\ \hline
126.gcc      & 10 & 2 & 15 & 15 &  8 &  8 \\ \hline
130.li       &  8 & 0 &  8 &  6 &  4 &  4 \\ \hline
132.ijpeg    & 14 & 3 & 17 & 17 & 15 & 15 \\ \hline
134.perl     & 11 & 1 & 18 & 17 & 12 & 12 \\ \hline
147.vortex   & 15 & 3 & 16 & 16 & 13 & 13 \\ \hline
\hline
\textbf{average} & 11 & 1 & 16 & 16 & 13 & 13 \\ \hline
\end{tabular}
\end{table}

Heuristic \Htwo\ does a significantly better job than
any of the other methods at limiting code growth.
This is not surprising, since it uses a simple, static threshold 
that only allows inlining of small procedures.
In general, heuristic \Hone, or \hankreg\ in its basic form,
does a little better in most cases than the remaining
heuristics based on demand-driven inlining.
The changes in code growth are generally slight, and in
nearly all cases remained under the static limit of 20\%
used in earlier experiments, indicating these heuristics
provide improved code growth control.
Code growth for \emph{compress} exceeded 20\% for 
\Hthree, \Hfour, \Hfive\ and \Hsix. 
This is due to an order of region formation, and therefore inlining,
that causes a very large procedure to be inlined when the code growth
limit had already nearly been reached.

\subsubsection*{Compilation time}

Results for compilation time for the heuristics
are presented in Table~\ref{tab:heuristics-comptime}.
The change in compilation time as compared with
\Hzero\ is shown as a percentage increase (positive) or
decrease (negative).  
Compilation time for each heuristic was measured by 
timing the compilation through the optimized \emph{Lcode} phase, 
just prior to the point when Trimaran outputs instrumented 
code for simulated execution on the target architecture.
This timing includes any applicable phases for profiling,
intermediate code generation, aggressive inlining, 
region formation (which may or may not include
demand-driven inlining), and region-based optimization.
The timings used were system times (i.e., wall clock
times) accurate to the nearest 10th of a seconds, and were
on the order of minutes or hours (not unusual for a
research compiler).

\begin{table}
\centering
\small
\caption{Percentage change in compilation time over \Hzero.}
\vspace{1ex}
\label{tab:heuristics-comptime}
\begin{tabular}{|l@{\hspace{6pt}}|r|r|r|r|r|r|} \hline
Benchmark    & \Hone & \Htwo & \Hthree & \Hfour & \Hfive &  \Hsix \\
\hline \hline
008.espresso &  2.1 &  0.0 &  2.0 &  4.2 &  4.9 &  4.5 \\ \hline
023.eqntott  &  1.8 & -0.1 &  2.4 &  4.8 &  9.3 &  7.9 \\ \hline
026.compress & -8.3 & -1.3 & -2.8 &  0.0 &  5.6 &  2.8 \\ \hline
099.go       &  3.0 & -2.5 &  6.5 &  7.8 & 10.0 &  9.3 \\ \hline
124.m88ksim  &  4.0 & -2.1 & 27.6 & 18.4 & 27.6 & 27.4 \\ \hline
126.gcc      &  2.9 & -0.3 & 21.1 & 15.2 & 15.4 & 15.2 \\ \hline
130.li       &  4.5 &  0.0 & 26.8 & 24.5 & 25.6 & 24.9 \\ \hline
132.ijpeg    &  3.5 & -1.0 & 13.9 & 13.5 & 14.0 & 13.8 \\ \hline
134.perl     &  2.8 & -1.4 & 14.8 & 14.2 & 15.3 & 15.1 \\ \hline
147.vortex   &  3.7 &  0.1 &  4.8 &  7.8 &  3.0 &  3.0 \\ \hline
\hline
\textbf{average} & 2.0 & -0.9 & 11.7 & 11.0 & 13.1 & 12.4 \\ \hline
\end{tabular}
\end{table}

In general, compilation time improves the most for 
\Htwo\ which uses the simplest inlining heuristic, 
and \Hone, the \hankreg\ compilation method.
Due to the overhead introduced in the current
implementation of demand-driven inlining, 
unusually high increases in compilation time
were seen in most other cases where demand-driven
inlining is used (\Hthree\ through \Hsix).

It is interesting to note that for some of the benchmarks,
\emph{008.espresso}, \emph{023.eqntott}, \emph{026.compress},
and \emph{147.vortex}, compilation time
increased only slightly over \Hone\ and \Htwo\
for the remaining heuristics.  This indicates that
other more complex factors may be helping to control compilation
time in spite of the more advanced and time-consuming 
inlining heuristics being used.

\subsubsection*{Runtime performance}

Relative changes in performance between the baseline 
heuristic \Hzero\ and the others are shown in
Table~\ref{tab:heuristics-speed}.
Performance was measured by running the programs
using the Trimaran simulator, which involved
an additional phase of compilation to instrument
the \emph{Lcode} output from region formation
to execute in the simulated \emph{EPIC} 
environment described earlier.

\begin{table}
\centering
\small
\caption{Percentage change in execution time over \Hzero.}
\vspace{1ex}
\label{tab:heuristics-speed}
\begin{tabular}{|l@{\hspace{6pt}}|r|r|r|r|r|r|} \hline
Benchmark    &  \Hone & \Htwo &  \Hthree &  \Hfour & \Hfive &  \Hsix \\
\hline \hline
008.espresso & -4.01 &  0.50 & -5.50 &  -7.35 &   1.75 &   1.75 \\ \hline
023.eqntott  & -3.17 &  1.31 & -4.02 &  -6.11 &  -1.90 &  -1.90 \\ \hline
026.compress & -3.11 &  0.00 & -3.11 &  -3.11 &  -2.98 &  -2.98 \\ \hline
099.go       & -3.19 &  0.07 &  2.30 &  -4.20 &  -4.10 &  -4.10 \\ \hline
124.m88ksim  & -6.13 & -1.31 & -3.90 &  -9.22 &  -9.13 &  -9.13 \\ \hline
126.gcc      & -5.20 & -1.03 & -4.15 &  -0.24 & -10.20 & -10.20 \\ \hline
130.li       & -8.49 & -2.16 & -4.01 & -12.53 & -12.20 & -12.20 \\ \hline
132.ijpeg    & -2.98 &  0.12 & -4.77 &  -7.33 &  -6.97 &  -6.97 \\ \hline
134.perl     & -5.50 & -1.90 & -4.79 & -10.43 &  -9.54 &  -9.54 \\ \hline
147.vortex   & -3.05 & -1.71 & -5.10 &  -9.25 &  -9.21 &  -9.21 \\ \hline
\hline
\textbf{average} & -4.48 & -0.61 & -3.71 & -6.98 & -6.45 & -6.45 \\ \hline
\end{tabular}
\end{table}

\Hthree\ was competitive with \Hone, but
\Hfour, \Hfive\ and \Hsix\ all showed general improvements
in performance over \Hone.
Overall, the best performance speedup was consistently demonstrated
with heuristic \Hfour, which uses the static loop call weight 
estimator and recursion prevention to guide inlining decisions.
%
%
%
%
%
There was also little or no significant change in
processor utilization (i.e., \emph{CPI}) for most of the benchmarks
under most of the heuristics.

\subsubsection{Discussion}

The code growth, compilation time and runtime performance
for the benchmarks under different
inlining heuristics interact in a number of ways.
For the cases that cause more code growth, execution time
also improves.  
The more naive heuristics of \Htwo\
lead to the smallest increases in code size and compilation time,
but also do not improve performance as much as the other more
sophisticated methods.  
Larger increases in code growth and
compilation time do not always translate to improvements in
execution time, indicating that
bounding code growth is indeed important, as was believed.
For example, when \Hthree\ was applied to \emph{099.go},
code size and compilation time increased more than
with \Htwo\, while execution took longer, possibly due to
recursive inlining of less important code.

The more scientific codes 
(\emph{124.m88ksim}, \emph{132.ijpeg}, \emph{147.vortex}),
tend to benefit the most from increases in code growth and
compilation time (which is also optimization and scheduling time)
in terms of their speedup, particularly
for the most advanced profile-estimating (\Hfour) and 
profile-based (\Hfive\ and \Hsix) methods.
Smaller benchmarks (\emph{026.compress}, \emph{023.eqntott}), by both size and 
number of procedures, are less predictable, although significant 
performance gains are seen with \Hthree\ through \Hsix, 
with most showing improvement over the original 
region-based technique (\Hone).
Benchmarks with more recursion 
(\emph{026.compress}, \emph{099.go}, \emph{130.li})
require more compilation time and gain
comparatively less in performance improvements than the others.

The combination of heuristics in \Hfour\ proved consistently
to be the most effective at controlling code growth and
compilation time while improving runtime performance.
The fact that \Hfour\ bases inlining decisions on the static
loop call weight, which estimates runtime behavior, rather
than the actual profiling information itself, 
as in \Hfive\ and \Hsix, is significant,
indicating that profiling may not be necessary for
making good demand-driven inlining decisions during region formation.
Profiling generally is more precise than static estimates
because it directly measures program behavior at runtime,
but requires more overhead and depends on the data used for the profiling.

\subsection{Impact on Compilation Unit Characteristics}

The experimental study in Section~\ref{section:demand} examined
how runtime performance can be improved by increasing 
interprocedural scope of compilation units, 
and reducing the profile variance of each unit.
To test this hypothesis further, the characteristics of
the compiled \emph{Lcode} were measured after
region formation for heuristics 
\Hzero\ (the baseline, with no inlining or region formation),
\Hone\ (\hankreg, for comparison) 
and \Hfour\ (the overall best performing heuristic).

\subsubsection*{Unit size}

Table~\ref{tab:4unitsize} compares the compilation unit
size characteristics resulting from the three heuristics,
\Hzero, \Hone, and \Hfour.
Both \Hone\ and \Hfour\ show significant
improvement in control of the size of the unit of
compilation, with average region sizes ranging from 3\% to 19\% of
the original average procedure sizes.
\Hfour\ consistently produces more compilation units than \Hone,
which is reflective of comparative code growth measurements
for the two heuristics.
The average size of the unit of compilation decreases slightly from
\Hone\ to \Hfour\ by 0.1 to 0.9 \emph{Lcode} instructions.
Although such a slight decrease in the average sizes of 
compilation units cannot directly account for the longer
compilation times seen with \Hfour,
the more significant increase factor is code growth
which results from the increase in the number of
compilation units that results from a decrease in
average size; with more code to compile, 
compilation time naturally is increased.

\begin{table}
\centering
\small
\caption{Comparison of number and average size of compilation units
for three heuristics.}
\vspace{1ex}
\label{tab:4unitsize}
\small
\begin{tabular}{|l||r|r||r|r||r|r|} \hline
	& \multicolumn{2}{c||}{\Hzero}
	& \multicolumn{2}{c||}{\Hone} 
	& \multicolumn{2}{c|}{\Hfour} \\
        & \multicolumn{1}{c|}{Avg} & \multicolumn{1}{c||}{Units} 
	& \multicolumn{1}{c|}{Avg} & \multicolumn{1}{c||}{Units} 
	& \multicolumn{1}{c|}{Avg} & \multicolumn{1}{c|}{Units} \\
Benchmark
	& \multicolumn{1}{c|}{size} & \multicolumn{1}{c||}{(proc.)} 
	& \multicolumn{1}{c|}{size} & \multicolumn{1}{c||}{(reg.)} 
	& \multicolumn{1}{c|}{size} & \multicolumn{1}{c|}{(reg.)} \\
\hline \hline
008.espresso & 183 &  361 & 15.9 &  4267 & 15.0 &  4710 \\ \hline
023.eqntott  & 169 &   62 & 16.6 &   656 & 16.2 &   755 \\ \hline
026.compress & 152 &   16 & 16.8 &   206 & 16.4 &   218 \\ \hline
099.go       & 234 &  383 & 17.0 &  5424 & 16.9 &  5921 \\ \hline
124.m88ksim  & 155 &  252 & 18.6 &  3494 & 17.7 &  4027 \\ \hline
126.gcc      & 530 & 1170 & 15.3 & 41613 & 14.9 & 43971 \\ \hline
130.li       &  81 &  357 & 14.9 &  2427 & 15.6 &  2516 \\ \hline
132.ijpeg    & 137 &  473 & 14.1 &  4720 & 13.8 &  5109 \\ \hline
134.perl     & 206 &  316 & 15.2 &  4395 & 14.9 &  4891 \\ \hline
147.vortex   & 161 & 1127 & 16.9 & 11026 & 16.2 & 11713 \\ \hline
\hline
\textbf{average} & 201 & 452 & 16.1 & 7823 & 15.8 & 8383 \\ \hline
\end{tabular}
\end{table}

\subsubsection*{Profile homogeneity}

Table~\ref{tab:4homogen} shows the profile homogeneity and
percentage of invariant code for these three heuristics.
In most cases, \Hfour\ improved upon the amount of
invariant code versus \Hone, while showing slight
to moderate increases in the profile variance.
The consistent increase in variance with the attending increase
in percentage of invariant compilation units indicates
that \Hfour, as compared with \Hone, 
is simultaneously improving the profile
homogeneity of more compilation units while increasing
the variance of a smaller number of compilation units
by relocating the more variant code.
The benefit seen to the percentage of invariant
compilation units reflects the improvement in runtime
performance for \Hfour\ over \Hone.

%
%
\begin{table}
\centering
\small
\caption{Comparison of percentage of invariant compilation units
and profile variance (homogeneity) for three heuristics, and
resulting interprocedural scope.}
\vspace{1ex}
\label{tab:4homogen}
\footnotesize
\begin{tabular}{|l@{\hspace{6pt}}||r|r||r|r||r|r||r|r|} \hline
 &
    \multicolumn{2}{c||}{\Hzero} &
    \multicolumn{2}{c||}{\Hone} &
    \multicolumn{2}{c||}{\Hfour} &
    \multicolumn{2}{c|}{Pct. interproc.}\\
&
    \multicolumn{1}{c|}{Profile} & \multicolumn{1}{c||}{Pct. units} &
    \multicolumn{1}{c|}{Profile} & \multicolumn{1}{c||}{Pct. units} &
    \multicolumn{1}{c|}{Profile} & \multicolumn{1}{c||}{Pct. units} &
    \multicolumn{2}{c|}{ops.} \\
Benchmark &
    \multicolumn{1}{c|}{variance} & \multicolumn{1}{c||}{invariant} &
    \multicolumn{1}{c|}{variance} & \multicolumn{1}{c||}{invariant} &
    \multicolumn{1}{c|}{variance} & \multicolumn{1}{c||}{invariant} &
    \multicolumn{1}{c|}{\Hone} & \multicolumn{1}{c|}{\Hfour} \\
\hline \hline
008.espresso & 0.368 & 93.1 & 0.288 & 93.7 & 0.361 & 94.0 & 18.0 & 19.2 \\ \hline
023.eqntott  & 0.020 & 97.7 & 0.001 & 97.8 & 0.022 & 98.6 & 15.1 & 23.8 \\ \hline
026.compress & 0.313 & 92.1 & 0.210 & 93.7 & 0.275 & 93.8 & 21.7 & 25.9 \\ \hline
099.go       & 0.372 & 91.5 & 0.310 & 91.9 & 0.331 & 93.1 & 23.1 & 24.9 \\ \hline
124.m88ksim  & 0.308 & 93.9 & 0.260 & 94.4 & 0.311 & 94.9 & 17.1 & 19.8 \\ \hline
126.gcc      & 0.323 & 92.9 & 0.262 & 93.4 & 0.300 & 93.9 & 18.2 & 20.4 \\ \hline
130.li       & 0.208 & 93.7 & 0.208 & 93.7 & 0.203 & 94.2 & 24.4 & 27.6 \\ \hline
132.ijpeg    & 0.281 & 93.2 & 0.201 & 93.8 & 0.239 & 94.2 & 17.3 & 19.3 \\ \hline
134.perl     & 0.270 & 94.3 & 0.189 & 94.5 & 0.249 & 95.2 & 16.9 & 18.1 \\ \hline
147.vortex   & 0.280 & 93.1 & 0.211 & 93.8 & 0.271 & 94.1 & 19.0 & 20.5 \\ \hline
\hline
\textbf{average} & 0.274 & 93.6 & 0.214 & 94.1 & 0.256 & 94.6 & 19.9 & 24.1 \\ \hline
\end{tabular}
\end{table}

\subsubsection*{Interprocedural scope}

Table~\ref{tab:4homogen} also compares the change in
interprocedural scope for \Hone\ and \Hfour\ compared
to the baseline heuristic \Hzero, which
had 0\% interprocedural code since no inlining
was performed.
Interprocedural scope improved in all cases
when using the demand-driven heuristics of
\Hfour, which showed improvements of
1.2\% to 8.7\% over \Hone, as well.
Improvements for \Hfour\ as compared with \Hone\ were less
significant for \emph{008.espresso}, \emph{099.go}, and \emph{134.perl},
which have more instances of direct recursion than
\emph{130.li}, which exhibits significant indirect recursion.
Indirect recursion within region formation leads to increased
interprocedural regions as procedures are inlined into other
procedures.  Direct recursion, or self-recursion, leads only to the
inlining of a procedure into itself, if at all.
The smaller size and lower number of procedures in \emph{023.eqntott}
and \emph{026.compress} led to the larger improvements to
interprocedural scope due to a higher proportion of smaller
procedures.

%
%

\section{Related Work}
\label{section:related}

Region-based compilation remains an active area of
research, with promising applications to a Java
virtual machine~\cite{Bala00,Suga03} including
a variety of adaptive techniques~\cite{Arno05},
and ILP optimization and scheduling frameworks~\cite{Liu03,Zhan04,Zhou01}.
Region formation is a form of interprocedural data
flow analysis, a well-researched area with many
benefits to ILP~\cite{Good97,Hall95,Reps95}.
Disadvantages are that during analysis it can have unscalable
memory requirements~\cite{Dues97}
or require exponential time with respect to
program size~\cite{Dues97}.
Advances address the issue of unscalable 
memory and time requirements by using 
modular~\cite{Lian99,Roun99} and
demand-driven~\cite{Dues97}
approaches,
while profile-driven analysis and 
optimization~\cite{Ammo98,Ayer97,Ball98,Bodi98,Cald97} 
are vital to code improvement and performance.

Procedure inlining is used to eliminate call 
overhead~\cite{Arno00,Baco94} leading to fewer 
and faster calls~\cite{Appe92},
improve compiler analysis and optimization~\cite{Appe92,Baco94},
register usage, code locality and execution speed~\cite{Baco94},
provide more precise data flow information to generate more efficient
code \emph{specialized} to the callee~\cite{Appe92,Arno00},
and enable intraprocedural analysis and optimizations 
such as constant propagation and elimination 
of redundant operations to be applied at
interprocedural scope~\cite{Baco94,Hall95}.
However, inlining can increase register 
pressure~\cite{Arno00,Baco94,Davi88},
code size~\cite{Arno00,Baco94},
instruction cache misses~\cite{Appe92,Arno00,Baco94},
and compilation time, which is more critical during dynamic 
compilation~\cite{Arno00}.
Extensive research into inlining heuristics and
the factors that bolster or limit their effectiveness
within procedure-oriented compilers has been 
performed~\cite{Alle88,Appe92,Ayer97,Baco94,
Chan92,Davi88,Hwu89,Jaga96,Kase98,Zhao03}.

\section{Conclusions and Future Work}
\label{section:conclusions}

Region-based compilation has already been shown to help increase
ILP performance by enabling interprocedural code motion
without the expense of large compilation units or
interprocedural data flow analysis.
This research has focused on improving the effectiveness of 
region-based compilation that integrates heuristics-guided
inlining into region formation analysis.
Experimental results comparing two region-based 
approaches demonstrated that a demand-driven approach to inlining,
as compared to a phased aggressive inlining approach,
can reduce memory requirements and code growth
while improving runtime performance due to
increased profile homogeneity and interprocedural scope.
These improvements are further enhanced by making
more informed inlining decisions, and reordering
the processing of procedures by a region-based compiler,
leading to further improvements to compilation unit 
characteristics, reflected as improved performance.
Heuristics based on static analysis can be as effective
as profile-based heuristics at guiding inlining decisions.  

Partial inlining is an inlining technique that selectively
inlines portions of a callee procedure into a callsite
rather than the entire body of the 
callee~\cite{Bala00,Coop99,Goub94,Monn00}.
Region-based compilation naturally enables a form
of partial inlining for the optimization phase of compilation~\cite{Hank97}.
The approach presented in this paper is being extended to the design of an
algorithm for incorporating partial inlining into
region-based compilation~\cite{Way02a}, including its applicability
to object-oriented programming.

\bibliographystyle{plain}
\bibliography{TechReport}

\begin{thebibliography}{10}

\bibitem{Aho88}
Alfred~V. Aho, Ravi Sethi, and Jeffrey~D. Ullman.
\newblock {\em Compilers: Principles, Techniques, and Tools}.
\newblock Addison-Wesley Publishing Company, 1988.

\bibitem{Alle88}
R.~Allen and S.~Johnson.
\newblock Compiling {C} for vectorization, parallelization, and inline
  expansion.
\newblock {\em ACM SIGPLAN Conference on Programming Language Design and
  Implementation}, pages 241--249, 1988.

\bibitem{Ammo98}
Glenn Ammons and James~R. Larus.
\newblock Improving data-flow analysis with path profiles.
\newblock In {\em ACM SIGPLAN Conference on Programming Language Design and
  Implementation}, pages 72--84, June 1998.

\bibitem{Appe92}
Andrew Appel.
\newblock {\em Compiling with Continuations}.
\newblock Cambridge University Press, 1992.

\bibitem{Arno05}
Matthew Arnold, Stephen Fink, David Grove, Michael Hind, and Peter~F. Sweeney.
\newblock A survey of adaptive optimization in virtual machines.
\newblock In {\em Proceedings of the {IEEE}}, pages 449--466, February 2005.

\bibitem{Arno00}
Matthew Arnold, Stephen Fink, Vivek Sarkar, and Peter~F. Sweeney.
\newblock A comparative study of static and dynamic heuristics for inlining.
\newblock In {\em {ACM SIGPLAN} Workshop on Dynamic and Adaptive Compilation
  and Optimization ({Dynamo 2000})}, pages 52--64, January 2000.

\bibitem{Ayer97}
Andrew Ayers, Richard Schooler, and Robert Gottlieb.
\newblock Aggressive inlining.
\newblock {\em ACM SIGPLAN Conference on Programming Language Design and
  Implementation}, pages 134--145, 1997.

\bibitem{Baco94}
David~F. Bacon, Susan~L. Graham, and Oliver~J. Sharp.
\newblock Compiler transformations for high-performance computing.
\newblock {\em Computing Systems}, 26(4):345--420, 1994.

\bibitem{Bala00}
Vasanth Bala, Evelyn Duesterwald, and Sanjeev Banerjia.
\newblock Dynamo: {A} transparent dynamic optimization system.
\newblock In {\em ACM SIGPLAN Conference on Programming Language Design and
  Implementation}, pages 1--12, Vancouver, Canada, June 2000.

\bibitem{Ball98}
Thomas Ball, Peter Mataga, and Mooly Sagiv.
\newblock Edge profiling versus path profiling: The showdown.
\newblock In {\em ACM SIGPLAN Symposium on the Principles of Programming
  Languages}, pages 134--148, 1998.

\bibitem{Bodi98}
Rastislav Bodik, Rajiv Gupta, and Mary~Lou Soffa.
\newblock Complete removal of redundant expressions.
\newblock In {\em ACM SIGPLAN Conference on Programming Language Design and
  Implementation}, pages 1--14, June 1998.

\bibitem{Cald97}
Brad Calder, Peter Feller, and Alan Eustace.
\newblock Value profiling.
\newblock In {\em {IEEE/ACM} International Symposium on Microarchitecture
  (MICRO)}, pages 259--269, December 1997.

\bibitem{Chan92}
Pohua~P. Chang, Scott~A. Mahlke, William~Y. Chen, and W.~W. Hwu.
\newblock Profile-guided automatic inline expansion for {C} programs.
\newblock {\em Software Practice and Experience}, 22(5):349--369, May 1992.

\bibitem{Coop99}
Keith~D. Cooper and Nathaniel McIntosh.
\newblock Enhanced code compression for embedded {RISC} processors.
\newblock In {\em ACM SIGPLAN Conference on Programming Language Design and
  Implementation}, May 1999.

\bibitem{Davi88}
J.~W. Davidson and A.~M. Holler.
\newblock A study of a {C} function inliner.
\newblock {\em Software Practice and Experience}, 18(8):775--790, August 1988.

\bibitem{Dues97}
Evelyn Duesterwald, Rajiv Gupta, and Mary~Lou Soffa.
\newblock A practical framework for demand-driven interprocedural data flow
  analysis.
\newblock {\em ACM Transactions on Programming Languages and Systems},
  19(6):992--1030, 1997.

\bibitem{Good97}
David~W. Goodwin.
\newblock Interprocedural dataflow analysis in an executable optimizer.
\newblock In {\em ACM SIGPLAN Conference on Programming Language Design and
  Implementation}, pages 122--133, June 1997.

\bibitem{Goub94}
Jean Goubault.
\newblock Generalized boxings, congruences and partial inlining.
\newblock In {\em 1st Static Analysis Symposium ({SAS '94})}, number 864 in
  Lecture Notes in Computer Science, pages 147--161. Springer Verlag, Namur,
  Belgium, September 1994.

\bibitem{Hall95}
Mary~W. Hall, Brian~R. Murphy, and Saman~P. Amarasinghe.
\newblock Interprocedural analysis for parallelization: Design and experience.
\newblock In {\em SIAM Conference on Parallel Processing for Scientific
  Computing}, pages 650--655, February 1995.

\bibitem{Hank97}
R.~E. Hank, W.~W. Hwu, and B.~R. Rau.
\newblock Region-based compilation: Introduction, motivation, and initial
  experience.
\newblock {\em International Journal of Parallel Programming}, 25(2):113--146,
  April 1997.

\bibitem{Hwu89}
W.~W. Hwu and P.~P. Chang.
\newblock Inline function expansion for compiling {C} programs.
\newblock {\em ACM SIGPLAN Conference on Programming Language Design and
  Implementation}, pages 246--257, 1989.

\bibitem{Hwu93}
W.~W. Hwu, S.~A. Mahlke, W.~Y. Chen, P.~P. Chang, N.~J. Warter, R.~A.
  Bringmann, R.~G. Ouellette, R.~E. Hank, T.~Kiyohara, G.~E. Haab, J.~G. Holm,
  and D.~M. Lavery.
\newblock The superblock: An effective structure for {VLIW} and superscalar
  compilation.
\newblock {\em Journal of Supercomputing}, 7:229--248, July 1993.

\bibitem{Jaga96}
Suresh Jagannathan and Andrew Wright.
\newblock Flow-directed inlining.
\newblock {\em ACM SIGPLAN Conference on Programming Language Design and
  Implementation}, pages 193--205, 1996.

\bibitem{Kase98}
O.~Kaser and C.~R. Ramakrishnan.
\newblock Evaluating inlining techniques.
\newblock {\em Computer Languages}, 24:55--72, 1998.

\bibitem{Lian99}
D.~Lian and M.~J. Harrold.
\newblock Efficient points-to analysis for whole-program analysis.
\newblock In {\em {ACM SIGSOFT} International Symposium on the Foundations of
  Software Engineering ({FSE})}, pages 199--215, September 1999.

\bibitem{Liu03}
Yang Liu, Zhaoquing Zhang, Ruliang Qiao, and Roy~Dz ching Ju.
\newblock A region-based compilation infrastructure.
\newblock In {\em Seventh Workshop on Interaction between Compilers and
  Computer Architectures}, pages 75--84, Anaheim, California, February 2003.

\bibitem{Monn00}
Stefan Monnier and Zhong Shao.
\newblock Inlining as staged computation.
\newblock Technical Report {TR}-1193, Department of Computer Science, Yale
  University, March 2000.

\bibitem{Trim98}
Amitabh Nene, Suren Talla, Benjamin Goldberg, and Rodric~M. Rabbah.
\newblock Trimaran - an infrastructure for compiler research in
  instruction-level parallelism - user manual, 1998.
  \verb+http://www.trimaran.org+.
\newblock New York University.

\bibitem{Reps95}
Thomas Reps, Susan Horowitz, and Mooly Sagiv.
\newblock Precise interprocedural dataflow analysis via graph reachability.
\newblock In {\em ACM SIGPLAN Symposium on the Principles of Programming
  Languages}, pages 49--61, San Francisco, California, January 1995.

\bibitem{Roun99}
Atanas Rountev, Barbara~G. Ryder, and William Landi.
\newblock Data-flow analysis of program fragments.
\newblock In {\em {ACM SIGSOFT} International Symposium on the Foundations of
  Software Engineering ({FSE})}, pages 235--252, September 1999.

\bibitem{Suga03}
Toshio Suganuma, Toshiaki Yasue, and Toshio Nakatani.
\newblock A region-based compilation technique for a java just-in-time
  compiler.
\newblock In {\em ACM SIGPLAN Conference on Programming Language Design and
  Implementation}, pages 312--323, June 2003.

\bibitem{Way01a}
Tom Way, Ben Breech, Wei Du, and Lori Pollock.
\newblock Demand-driven inlining heuristics in a region-based optimizing
  compiler for {ILP} architectures.
\newblock In {\em International Conference on Parallel and Distributed
  Computing and Systems}, pages 90--95, Anaheim, California, 2001.

\bibitem{Way00}
Tom Way, Ben Breech, and Lori Pollock.
\newblock Region formation analysis with demand-driven inlining for
  region-based optimization.
\newblock In {\em International Conference on Parallel Architectures and
  Compilation Techniques (PACT)}, pages 24--33, Philadelphia, Pennsylvania,
  October 2000.

\bibitem{Way02a}
Tom Way and Lori Pollock.
\newblock A region-based partial inlining algorithm for an {ILP} optimizing
  compiler.
\newblock In {\em International Conference on Parallel and Distributed
  Processing Techniques and Applications({PDPTA})}, pages 552--556, Las Vegas,
  Nevada, 2002.

\bibitem{Zhan04}
Ziangyu Zhang and Rajiv Gupta.
\newblock Whole execution traces.
\newblock In {\em {IEEE/ACM} International Symposium on Microarchitecture
  (MICRO)}, pages 105--116, Portland, Oregon, December 2004.

\bibitem{Zhao03}
Peng Zhao and Jos\'{e}~Nelson Amaral.
\newblock To inline or not to inline?
\newblock In {\em 16th Workshop on Languages and Compilers for Parallel
  Computing {LCPC}}, pages 405--419, College Station, Texas, October 2003.

\bibitem{Zhou01}
Huiyang Zhou, Matthew~D. Jennings, and Thomas~M. Conte.
\newblock Tree traversal scheduling: A global scheduling technique for
  {VLIW}/{EPIC} processors.
\newblock In {\em 14th Workshop on Languages and Compilers for Parallel
  Computing {LCPC}}, pages 223--238, Cumberland Falls, Kentucky, August 2001.

\end{thebibliography}

\end{document}